\documentstyle[mncite]{mn}
\input{epsf}
\newcommand{\lqt}{\stackrel{<}{_{\sim}}}

\def\spose#1{\hbox to 0pt{#1\hss}}
\def\simlt{\mathrel{\spose{\lower 3pt\hbox{$\mathchar"218$}}
     \raise 2.0pt\hbox{$\mathchar"13C$}}}
\def\simgt{\mathrel{\spose{\lower 3pt\hbox{$\mathchar"218$}}
     \raise 2.0pt\hbox{$\mathchar"13E$}}}
\def\refindent{\par\noindent\hangindent=3pc\hangafter=1 }

\def\et{{\it et~al. }}
\def\hmpc{\;h^{-1}{\rm Mpc}}
\def\h-1mpc{\;h\;Mpc^{-1}}
\def\h2mpc{\;h^{-2}{\rm Mpc}}

\def\kms{{\rm km\;s}^{-1}}

\def\bj{b_{\rm J}}

\begin{document}
\title[Power Spectrum of STROMLO-APM]{
Power Spectrum Analysis of the Stromlo-APM Redshift Survey}
\author[H. Tadros and G.
Efstathiou]{H. Tadros\thanks{Current address: Astronomy Centre,
University of Sussex, Falmer, Brighton, UK} and G.
Efstathiou \\
Department of Physics, University of Oxford, Keble Road, Oxford, OX1
3RH. UK. }

\maketitle
\begin{abstract}

We test estimators of the galaxy power spectrum $P(k)$ against
simulated galaxy catalogues constructed from N-body simulations and
we derive formulae to correct for biases. These estimators are then
applied to compute the power spectrum of galaxies in the Stromlo-APM
redshift survey. We test whether the amplitude of $P(k)$ depends on
galaxy luminosity, but find no significant luminosity dependence
except at absolute magnitudes brighter than $M_{\bj} = -20.3$, ($H_{0} =
100 \kms$) where there is some evidence for a rise in the amplitude of
$P(k)$. By comparing the redshift space power spectrum of the
Stromlo-APM survey with the real space power spectrum determined from
the parent APM Galaxy Survey, we attempt to measure the distortion in
the shape of $P(k)$ caused by galaxy peculiar motions. We find some
evidence for an effect, but the errors are large and do not exclude a
value of $\beta = \Omega^{0.6}/b = 1$, where $\Omega$ is the
cosmological density parameter and $b$ is the linear biasing parameter
relating galaxy fluctuations to those in the mass, $\left(\delta
\rho/\rho\right)_{gal} = b \left(\delta \rho/\rho\right)_{m}$. The
shape of the Stromlo-APM power spectrum is consistent with that
determined from the CfA-2 survey, but has a slightly higher amplitude
by a factor of about 1.4 than the power spectrum of IRAS galaxies.

\end{abstract}
\section{Introduction}

The Stromlo-APM redshift survey is a $1$ in $20$ sparsely sampled
subset of $1787$ galaxies selected from the APM Galaxy survey (Maddox
\et 1990a) to a magnitude limit of $\bj=17.15$. The survey has been
used in several investigations of the large-scale clustering of
galaxies: Loveday \et (1992a) have investigated galaxy counts in nearly
cubical cells with sizes in the range $5 \hmpc$ -- $60 \hmpc$; Loveday
\et (1995a) have computed the two-point galaxy correlation function of
the Stromlo-APM survey and Loveday \et (1995b) have investigated
anisotropies in the two-point correlation function caused by
redshift-space distortions (see Kaiser 1987). In this paper, we
present an analysis of the power spectrum, $P(k)$, of galaxies in the
Stromlo-APM survey. The power spectrum of the galaxy distribution has
been estimated from a number of other surveys (see Efstathiou 1996
for a review).  Power spectra for optically selected samples have been
computed by {\it e.g.} Baumgart and Fry (1991), Vogeley \et (1992),
Park \et (1994) and Baugh and Efstathiou (1993, 1994). Power spectra
for various IRAS redshift surveys have been compute by 
Fisher \et (1993), Feldman, Kaiser and Peacock (1994, hereafter
FKP) and Tadros and Efstathiou (1995).

The layout of this paper is as follows. The Stromlo-APM redshift
survey is described in Section 2. In Sections 3 and 4, we discuss
estimators of the power spectrum and their biases and test our
techniques against simulated Stromlo-APM redshift surveys constructed
from large N-body simulations. In Section 5 we present estimates of
the power spectrum for volume limited and flux limited samples
selected from the Stromlo-APM survey and we investigate the
sensitivity of the results to the volume limit and the weights
assigned to the galaxies. A number of authors have claimed that
luminous galaxies are more strongly clustered than less luminous
galaxies ({\it e.g.} Hamilton 1988, Santiago and da Costa 1990, Iovino
\et 1993, Park \et 1994). We investigate the dependence of clustering
strength with luminosity in Section 5. In Section 6 we investigate the
distortion of the power spectrum measured in redshift space caused by
galaxy peculiar motions and we attempt to measure the effect by
comparing the redshift-space estimates of $P(k)$ from the Stromlo-APM
survey with real-space estimates of $P(k)$ inferred from the parent
APM catalogue (see Baugh and Efstathiou 1993). We present our main
conclusions in Section 7 and compare our estimates of $P(k)$ with
those for the CfA-2 survey and IRAS surveys and with the predictions
of various cold dark matter models.

\section{Data Set}
The Stromlo-APM redshift survey is described in detail by Loveday
\et (1992a).  The survey covers an area approximately defined
by the equatorial coordinates $21^{h}\lqt\alpha\lqt5^{h},
-72.5^{\circ}\lqt\delta\lqt-17.5^{\circ}$. In the analysis described
below, we analyse flux limited and
volume limited samples. The volume limited samples are 
constructed by removing all galaxies with redshifts
$z > z_{max}$ and absolute magnitudes  $M>M_{crit}$,
where 
\begin{equation}
M_{crit}=m_{lim}-25-5log\left(d_L(z_{max})\right)-kz_{max},
\vspace{0.5cm}
\end{equation}
$m_{lim}=17.15$ is the magnitude limit of the
survey  and $d_L(z_{max})$ is the luminosity distance
at redshift $z_{max}$. Throughout this paper, we assume
a spatially flat universe with $\Omega=1$, thus
\begin{equation}
d_L(z)=\frac{2c}{H_{0}}\left[1-\frac{1}{\sqrt{1+z}}\right](1+z).
\end{equation}
The median redshift of the Stromlo-APM survey is $z = 0.05$,
hence the results presented below are insensitive to the
assumed cosmological model. We adopt a $k$ correction 
of $3z$ in equation (1) which is which is appropriate for 
the median morphological type in the Stromlo-APM survey
in the $b_J$ passband (Efstathiou, Ellis and Peterson 1988).

There are some advantages in estimating power spectra from volume
limited rather than flux limited samples. Firstly, every galaxy in a
volume limited sample carries equal weight, whereas weighting factors
that are a function of the power spectrum $P(k)$ are required to
determine a minimum variance estimate of $P(k)$ from a flux limited
sample if the underlying density field is assumed to be Gaussian (see
FKP and Section 3).  Secondly, the analysis of flux limited samples
requires a model for the mean galaxy density $\overline n(r)$ as a
function of radial distance.  Thirdly, by estimating power spectra of
progressively larger volume limited samples it is possible, given a
large enough sample, to test whether the clustering amplitude is a
function of luminosity. These points will be discussed in further
detail in Section 4.

\begin{figure}
\centering
\begin{picture}(200,200)
\includegraphics{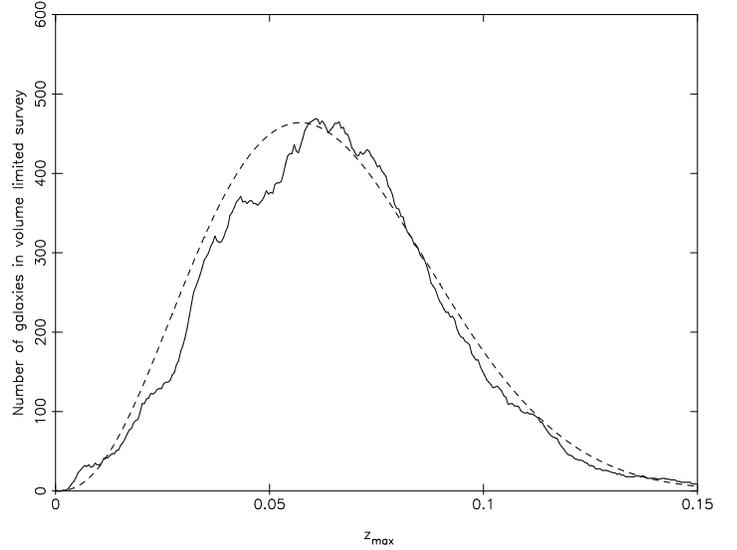}
\end{picture}
\caption{\label{mcritvz} The solid line shows the 
number of galaxies in volume limited subsamples of the 
Stromlo-APM survey as a 
function of limiting redshift $z_{max}$. The dashed  curve shows the
expected distribution derived by integrating the luminosity
function from Loveday $\et$ (1992a)  using the 
Schechter function parameters given in the text.}
\end{figure}

The solid line in Figure~\ref{mcritvz} shows the 
number of galaxies in volume limited subsamples of
the Stromlo-APM survey as a function of the limiting
redshift $z_{max}$. The total number of galaxies
peaks  at $z_{max}=0.06$, corresponding to an absolute
magnitude limit of $-19.34$\footnote{Throughout this
paper, we write the Hubble constant as
$H_0 = 100h\;{\rm km}\;{\rm s}^{-1}\;{\rm Mpc}^{-1}$
and quote absolute magnitudes assuming $h=1$ unless otherwise
stated.}. For most of the analysis in 
this paper we adopt a volume limit of $z_{max} =0.06$
(coordinate distance of $x_{max} = 172 \hmpc$)
containing $469$ galaxies. The dashed curve in  Figure~\ref{mcritvz}
shows the number of galaxies  predicted by integrating
over the galaxy luminosity function:
\begin{equation}
N\left(z_{max}\right) =  \phi_{\star} V\left(z_{max}\right)
\int_{L_{min}(z_{max})}^{\infty}\hspace*{-0.3cm}
\left(\frac{L}{L_{\star}}\right)^{\alpha}e\left(-\frac{L}{L_{\star}}\right)
d\left( \frac{L}{L_{\star}}\right)
\end{equation}
where $V\left(z_{max}\right)$ is the comoving volume 
of the survey to redshift $z_{max}$, and $L_{min}$ is the 
luminosity of a  galaxy with absolute magnitude
$M_{crit}$ (equation 1). We have used the luminosity function
parameters from Loveday \et (1992b) 
$\phi_{\star}$ = $1.12 \times 10^{-2}\;h^3{\rm Mpc}^{-3}$, 
$\alpha = -1.11$ and $M_{\star} = -19.73$; these are the best
fitting Schechter (1976) function parameters for galaxies with
absolute magnitudes in the range $-22 < M < -15$, uncorrected
for magnitude errors.

\section{Measurement of the Power Spectrum}

In this Section we establish our notation and 
describe our estimator of $P(k)$.
We imagine that the Universe is divided into
infinitesimal cells of volume $\delta V$ containing $n_i$ galaxies
such that the $n_i$ are either $0$ or $1$. The mean and variance
of $n_i$ are therefore,
\begin{eqnarray}
\left \langle n_i \right \rangle   =  \overline n \delta V \\
\left \langle n_i^2 \right \rangle  =   \left \langle n_i \right
\rangle
\end{eqnarray}
The observed galaxy count $n_o({\bf x}_i)$ at point 
${\bf x}_i$ is related to the real
galaxy count $n_i$ by
\begin{eqnarray}
 n_o({\bf x}_i)   =  n_i W({\bf x}_i),
\end{eqnarray} 
where $W({\bf x}_i)$ defines the window function of the 
survey. In general, the window function depends on the radial
distribution of galaxies in the survey, the survey boundaries
on the celestial sphere, and the weights applied to each galaxy.
In this section, we assume that we are analysing a volume limited
sample in which each galaxy is assigned equal weight. In this
case, the window function $W({\bf x})$ is equal to unity or zero 
according to whether a particular patch of the universe is included or 
excluded from the survey. It is straightforward to generalize the equations in 
this Section to  more complex window functions and to 
include radially dependent galaxy weights (see FKP).

For the Stromlo-APM survey, the window function is defined by an
angular mask consisting of the APM Galaxy Survey boundaries
together with rectangular holes around regions on
the plates where there are bright stars, satellite trails,
globular clusters, step-wedges {\it etc} (see Plate 2 of Maddox
\et 1990b). We estimate the Fourier transform of the window
function 
\begin{equation}
\hat W({\bf k}) = \frac{1}{V}\int W({\bf x})e^{i{\bf k  \cdot
x}}\,\mbox{d}^3{\bf x}
\end{equation}
by generating a large number random points with mean
density $\overline{n}_R$ within the survey 
volume and computing 
\begin{equation}
\hat W_e({\bf k}) = \frac{1}{\overline{n}_R V}\sum_{i} e^{i{\bf k  \cdot
x}_i},
\end{equation}
where the sum extends over all random points located at positions
${\bf x}_i$. The volume $V$ is a normalizing volume which  encloses
the survey.

We compute the Fourier transform of the observed density field
\begin{equation}
\hat n_o ({\bf k})  =  \frac{1}{V} \sum_i n_i e^{i {\bf k \cdot x}_i},
\end{equation}
and we define a variable with zero mean 
\begin{equation}
\Delta ({\bf k})  =  \hat n_o ({\bf k}) - \overline{n} \hat W({\bf
k}),
\end{equation}
where $\overline n$ is the mean galaxy density. The variance of
$\Delta ({\bf k})$ is related to the power spectrum $P(k)$ of
the galaxy distribution according to 
\begin{equation}
\langle \vert \Delta \left({\bf k}\right) \vert^2 \rangle
 = \frac{\overline n}{V} \sum_{\bf k^\prime} \vert \hat
W ({\bf k^\prime}) \vert^2 + \frac{\overline n^2}{V}
\sum_{\bf k^\prime} \vert \hat
W ({\bf k} - {\bf k^\prime}) \vert^2 P({\bf k^\prime}).
\end{equation}

\begin{figure}
\centering
\begin{picture}(200,200)
\includegraphics{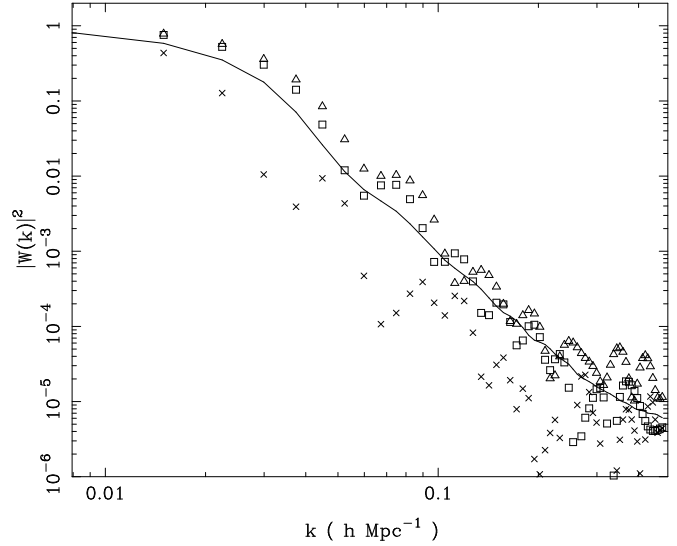}
\end{picture}
\caption{\label{plotwindow} The Fourier transform of
the Stromlo-APM survey volume limited at $z = 0.06$.
The solid line shows the window function averaged in radial
shells in ${\bf k}$-space while the
crosses, squares and triangles show $\vert W(k_x,0,0)\vert^2$, 
$\vert W(0,k_y,0)\vert^2$, and
$\vert W(0,0,k_z)\vert^2$ components respectively.
The x-axis is aligned approximately along the radial direction 
pointing to the centre of the survey on the celestial sphere.
We have normalized the curves so that they are equal to unity
at $k=0$.
}
\end{figure}

Figure~\ref{plotwindow} shows the window function of the Stromlo-
APM survey region volume limited at $z = 0.06$. This figure shows 
that the window function is sharply peaked in ${\bf k}$-space, falling off
approximately as $k^{-4}$. We have computed the sum in equation
(8) by Fast Fourier transforming the random density field
within a cubical volume of side $840 \hmpc$. We also plot
the components of $\vert W({\bf k}) \vert^{2}$ along the principal
axes of the cube, showing that the window function is anisotropic
and is narrower in the $x$-direction, which is aligned along
the radial direction of the centre of the survey.

Since the window function is sharply peaked in ${\bf k}$-space, it
is a good approximation to remove $P({\bf k^\prime})$ from the
summation in equation (11), thus our estimate of $P({\bf k})$ is
\begin{equation}
P_e({\bf k}) 
 = \frac{ \frac{V^2}{N_G^2} \vert \Delta({\bf k}) \vert^2
- \frac{1}{N_G} - \frac{1}{N_R}}{ \frac{V}{V_S}\sum_{\bf k^\prime}
\vert W_e({\bf k^\prime}) \vert^2},
\end{equation}
where $N_G$ and $N_R$ are the total number of galaxy and
random points within the survey volume $V_s$,
\begin{equation}
V_s = \int W({\bf x}) \;d^3{\bf x}.
\end{equation}
The negative terms in equation (12)
correct for Poisson shot noise in the galaxy and random
number distributions. Notice that in equation (12) we have
assumed that the mean galaxy density is equal to $N_G/V_s$.
The estimate $P_e$ will therefore be biased low because
the estimate of the mean galaxy density is affected by
galaxy clustering. An expression for this bias is derived
in Appendix A.

We estimate the sampling errors on the power spectrum in
two different ways. The first method is based on the
error analysis of FKP who assume that the galaxy density
field is a Gaussian point process. In the case of a volume limited 
survey, equation (2.4.6) of FKP for the
error on the point with wave vector $k$ is, in our notation,
\begin{eqnarray*}
\sigma_P(k) = \frac{V_s}{N_{G}}
\frac{(1+ \frac{N_G}{N_R}+\overline{n}P(k))}{\sum_{{\bf k^\prime}} \left|\hat{W}({\bf
k^\prime})\right|^2}
\end{eqnarray*}
\begin{equation}
\hspace*{3cm}\times\left(\frac{1}{N_{sum}} 
\sum_{\bf{k}^{\prime}}\sum_{\bf{k}^{\prime \prime}}\left|\hat{W}({\bf
k}^{\prime} - {\bf{k}}^{\prime \prime})\right|^2\right)^{\frac{1}{2}}
\end{equation}
In equation (14), $\bf{k}^{\prime}$ and $\bf{k}^{\prime \prime}$
are constrained to lie in the same shell with modulus $k$ and 
$N_{sum}$ is the total number of terms in the double summation.

The second method is based on the variance of $P(k)$ 
derived from
mock Stromlo-APM surveys constructed from N-body simulations. The mock
surveys, and a comparison with the FKP errors derived from equation (14)
are described in Section 4.2. This is a useful test because equation
(14) assumes Gaussian statistics, whereas the true galaxy density field
is non-Gaussian at least on scales $\simlt 5 \hmpc$ where the 
{\it rms} fluctuations exceed unity. By comparing the errors
from mock catalogues  which are non-Gaussian on small scales,
we can check the accuracy of  equation (14) over the full
range of wavenumbers presented in this paper.

In the above analysis, we have assumed that we are analysing
volume limited samples and assigning an equal weight to
each galaxy. The analysis of a flux limited sample is
more complicated because the window function $W({\bf x})$ 
depends on radial distance, and because we must assign
radially dependent weights to the galaxies to minimise the
variance on $P(k)$. These points are discussed in detail by
FKP and we follow their analysis in this paper. If the
underlying density fluctuations are Gaussian, FKP show that
the variance on the estimated $P(k)$ is minimised if each galaxy 
is assigned a weight 
\begin{equation}
w(r) = \frac{1}{1 + \overline{n}(r)P(k)},	
\end{equation}
where $\overline n (r)$ is the mean galaxy density as
a function of radial distance $r$, which we compute from
the luminosity function of the Stromlo-APM survey 
with the parameters given in Section 2. From equation
(15), we see that the minimum variance weighting for
each wavenumber $k$, depends on the true value of the
power spectrum at wavenumber $k$. Rather than applying
different weights at each wavenumber, we have estimated
the power spectrum from the flux-limited survey for
four  values of $P(k)$ in equation (15), 
$P\left(k\right)$ =  4000, 8000, 16000 and 32000
$\left(\hmpc\right)^{3}$, which span the range of
interest at wavenumbers $\simlt 0.3 h{\rm Mpc}^{-1}$.
Our analysis of flux limited surveys follows that of FKP except that
we compute $\alpha$ (the ratio of the space densities
in the real catalogue to that in the random catalogue, equation 2.1.3
of FKP) from the ratio of the sums
\begin{equation}
\sum_i {1 \over (1 + 4 \pi \overline n(r_{i})J_3)} \; \; ,
\end{equation}
instead of $\alpha = N_G / N_R$, 
where the summations are over all galaxies and random 
points. We have set $4 \pi J_3 = 10000
(h^{-1} {\rm Mpc})^3$  (see Tadros and Efstathiou 1995).

\section{Tests of the estimators using N-body
simulations}

In this Section, we investigate the accuracy with which we
can recover the power spectrum from the Stromlo-APM survey
using the methods described in the previous Section. There
are three key aspects of the analysis that we wish to test:
(i) the assumption that the convolution of the window function
with the power spectrum in equation (11) can be replaced
by a product as in equation (12); (ii) the bias
in the estimate of $P(k)$  caused by estimating the mean
galaxy density from the survey itself (Appendix A); (iii)
the accuracy of the FKP error estimates (equation 14).
We test these points by analysing mock Stromlo-APM surveys
constructed from N-body simulations.

\subsection{Numerical Simulations}
The numerical simulations that we use here are  described
by Croft and Efstathiou (1994). They consist of three ensembles of
10 simulations each containing 
$10^{6}$ particles within a periodic computational box
of length $\ell = 300 \hmpc$. The simulations were run with the
particle-particle-particle-mesh (P$^{3}$M) code described by
Efstathiou \et (1985). The simulations model  gravitational
clustering in a cold dark matter (CDM) dominated universe
with scale invariant initial density fluctuations.
The three ensembles are as follows: the standard CDM model
(Davis \et 1985), {\it i.e.} a spatially flat universe
with  $\Omega_{0}=1$ and $h=0.5$ (SCDM); a spatially flat
low density CDM universe with $\Omega_{0}=0.2$ 
and  a cosmological constant
$\lambda=\frac{\Lambda}{3H_{0}^{2}}=\left(1-\Omega_{0}\right)=0.8 $
(LCDM); a spatially flat mixed dark matter model in which CDM
contributes $\Omega_{CDM}=0.6$, baryons contribute
$\Omega_b = 0.1$ and massive neutrinos contribute $\Omega_\nu = 0.3$.

For the SCDM and LCDM simulation, the initial power spectrum 
is given by 
\begin{equation}
P\left(k\right)\propto
\frac{k}{\left[1+\left(ak+ \left(bk\right)^{\frac{3}{2}}+
\left(ck\right)^{2} \right)^{\nu}\right]^{\frac{2}{\nu}}}
\end{equation}
where $\nu=1.13$,
$a=\frac{6.4}{\Gamma}\hmpc$,$b=\frac{3.0}{\Gamma}\hmpc$ and
$c=\frac{1.7}{\Gamma}\hmpc$. Equation (17) is a good
approximation to the linear power spectrum of scale-invariant CDM
models with low baryon density, $\Omega_{b}\ll \Omega_{0}$
(Bond and Efstathiou 1984). The parameter $\Gamma$ in equation
(17) is equal to $\Omega_0h$, thus $\Gamma = 0.5$ for the
SCDM ensemble and $\Gamma = 0.2$ for the LCDM ensemble.
The initial conditions for our MDM simulations are generated
from the power spectrum given by equation 1 of Klypin \et
(1993). In the MDM models, we ignore the thermal motions
of the neutrinos and so follow the evolution of a collisionless
cold component with $\Omega_0 = 1$.

The final output times of the models are chosen to approximately
match the microwave background anisotropies measured in the
first year COBE maps (Smoot \et 1992) ignoring any contribution
from gravitational waves. Thus the {\it rms} mass fluctuations in
spheres of radius $8 \hmpc$ are $\sigma_8 = 1$ for the SCDM
and LCDM ensembles and $\sigma_8 = 0.67$ for the MDM ensemble.
Normalizing the anisotropies measured from the combined
first and second year COBE maps would increase
these values of $\sigma_8$ by $\sim 20$--$30\%$, but these
small changes are unimportant for most of the discussion below.

\subsection{Mock Stromlo-APM Surveys}

From the simulations described above we have constructed mock
Stromlo-APM surveys. We assume that galaxies are distributed like the
mass points in the simulations and make no attempt to introduce
biasing. We apply the APM angular mask to the simulations and
replicate the periodic box where necessary to generate distant points.
We then generate fully sampled volume limited catalogues to $z_{max} =
0.06$ that contain typically $100,000$ mass points. We also generate
volume limited samples containing on average $469$ points by random
sampling the mass points. These mock surveys thus have similar numbers
of points as there are galaxies in the Stromlo-APM survey at this
volume limit. Flux limited surveys are also generated by selecting
mass points with the radial selection function of the Stromlo-APM survey 
(see Section 2).

\begin{figure}
\centering
\begin{picture}(300,350)
\includegraphics{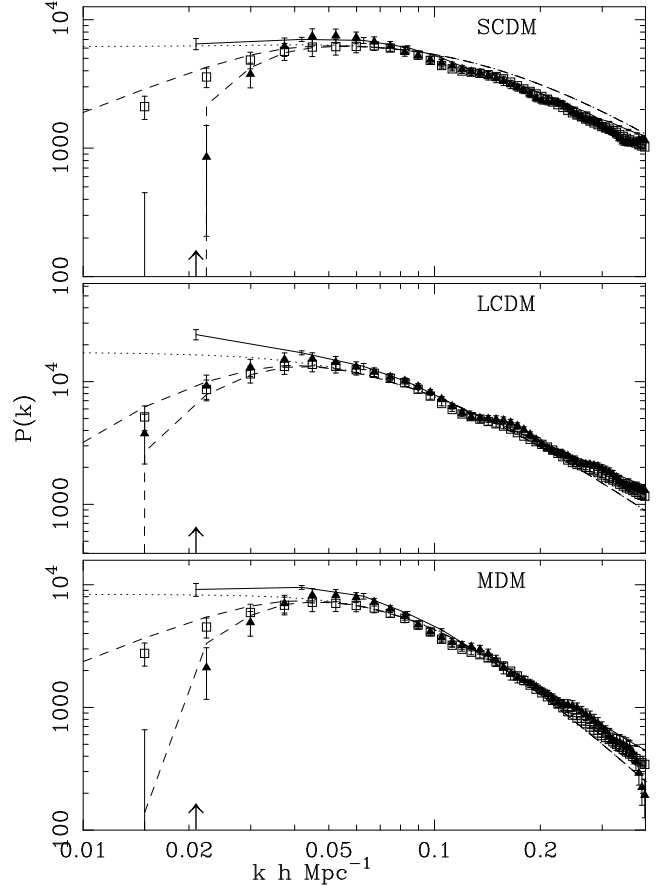}
\end{picture}
\caption{\label{plotpsrectest} Estimates of the power spectra
in real space for the three ensembles of N-body simulations described
in the text. The solid lines show the average of the power spectra
derived from the full cubical volume of the simulations. The open
squares show the average of $P(k)$ derived from equation (12) applied
to $10$ fully sampled realizations of the Stromlo-APM survey, volume
limited at $z_{max} = 0.06$. The filled triangles show the average of
$P(k)$ over $30$ realizations of the sparse-sampled Stromlo -APM
survey volume limited at $z_{max} = 0.06$. The error bars in each case
show the standard deviation of the mean. The dashed lines are derived
from equation (A4.2) of Appendix A, assuming the linear theory form of
the power spectrum for each ensemble and  illustrate how the mock
surveys are affected by the bias associated with the estimate of the
mean galaxy density. The dotted lines show the convolution of the
linear power spectrum (equation 17) with the window function for the
Stromlo-APM survey. The arrows show the smallest wavenumber
represented in the simulations, $k = 2 \pi/\ell_b$, where $\ell_b =
300\hmpc$ is the computational box length.}
\end{figure}

Figure~\ref{plotpsrectest} shows three estimates
of the power spectrum in real space measured for
each ensemble. The solid lines show the mean of the power 
spectra determined from the full cubical N-body simulations. 
These are computed by Fast Fourier transforming the particle distribution
on a $128^3$ grid using the nearest grid point assignment
scheme, as described by Efstathiou \et (1985).
The open squares show the power spectra derived from
equation (12) for the fully sampled mock Stromlo-APM surveys
volume limited at $z_{max} = 0.06$. The symbols show the mean
value of $P(k)$ from the ten members of each ensemble and the
error bars show the standard deviation on the mean derived
from the scatter in the $P(k)$ estimates. The open squares
are in very good agreement with the power spectra of the
full simulations at wavenumbers $k \simgt 0.04 h{\rm Mpc}^{-1}$,
showing that replacing the convolution over $P(k)$ in equation 
(11) by a product is an excellent approximation. At smaller
wavenumbers, the power spectrum is systematically underestimated.
This is caused by the bias associated with the mean density estimate
described in Appendix A. The dashed curves in Figure 3 show the
linear theory power spectra for each model including an
approximate  correction for the bias derived from equation
(A4.2) of Appendix A. These curves are in excellent agreement
with the data points, showing that the bias can be removed if
the form of the power spectrum is known. The filled triangles
show the power spectra for volume limited ($z_{max} = 0.06$)
realizations of the sparse sampled Stromlo-APM survey. As described
above, these simulations have approximately  the same number of 
objects as in the real catalogue and accurately model the window
function of the real survey. Because the number of points in
each realization is small (and hence the errors on $P(k)$ are
large), we have plotted the averages over $30$ realizations 
from each ensemble. The results are very similar to those
from the fully sampled Stromlo-APM simulations, except that
the bias associated with the mean density estimate is larger,
as expected from equation (A4.2) of Appendix A (shown by the
dashed lines passing through the points).

In Figure 4 we analyse flux limited sparse-sampled Stromlo-APM surveys
for the SCDM and LCDM ensembles using the techniques of FKP. Each
panel shows the power spectra averaged over $10$ realizations for four
values of $P(k)$ in the weighting scheme of equation (15) as indicated
in each panel. The results are qualitatively similar to those of
Figure 3 and show that the FKP estimator provides an unbiased estimate
of $P(k)$ at wavenumbers $k \simgt 0.04 h{\rm Mpc}^{-1}$. Furthermore,
Figure 4 shows that the estimates of $P(k)$ and the errors on $P(k)$
are relatively insensitive to the weighting scheme. This is as
expected since a minimum variance estimator should not be sensitive to
small departures from the minimum variance weighting scheme. The
errors on $P(k)$ at wavenumbers $k \simgt 0.1 h{\rm Mpc}^{-1}$ are
noticeably larger if we adopt a constant value of $P(k) \ge 16000
(h^{-3}{\rm Mpc})^3$ in the weighting scheme (15), but it is clear
that this is an overestimate of the amplitude of the power spectrum at
these wavenumbers.

We have also investigated the accuracy of the error estimates derived
from equation (14) and its generalization to flux limited
samples by comparing with the dispersion in the $P(k)$
estimates from our mock surveys. Equation (14) provides an accurate
estimate of the errors for wavenumbers $k\lqt 0.1 h {\rm Mpc}^{-1}$,
but tends to underestimate the errors by a factor of $\sim 2$ at
larger wavenumbers.

In summary, the results of this Section show  that our
methods provide nearly unbiased estimates of the power spectrum 
at wavenumbers $k \simgt 0.04 h{\rm Mpc}^{-1}$. At smaller
wavenumbers, the estimates are biased low because we determine
the mean galaxy density from the sample itself. The power
spectra derived from flux limited samples are insensitive to
the weighting scheme, provided we adopt a reasonable value
of $P(k)$ in equation (15).

\begin{figure}
\centering
\begin{picture}(200,500)
\includegraphics{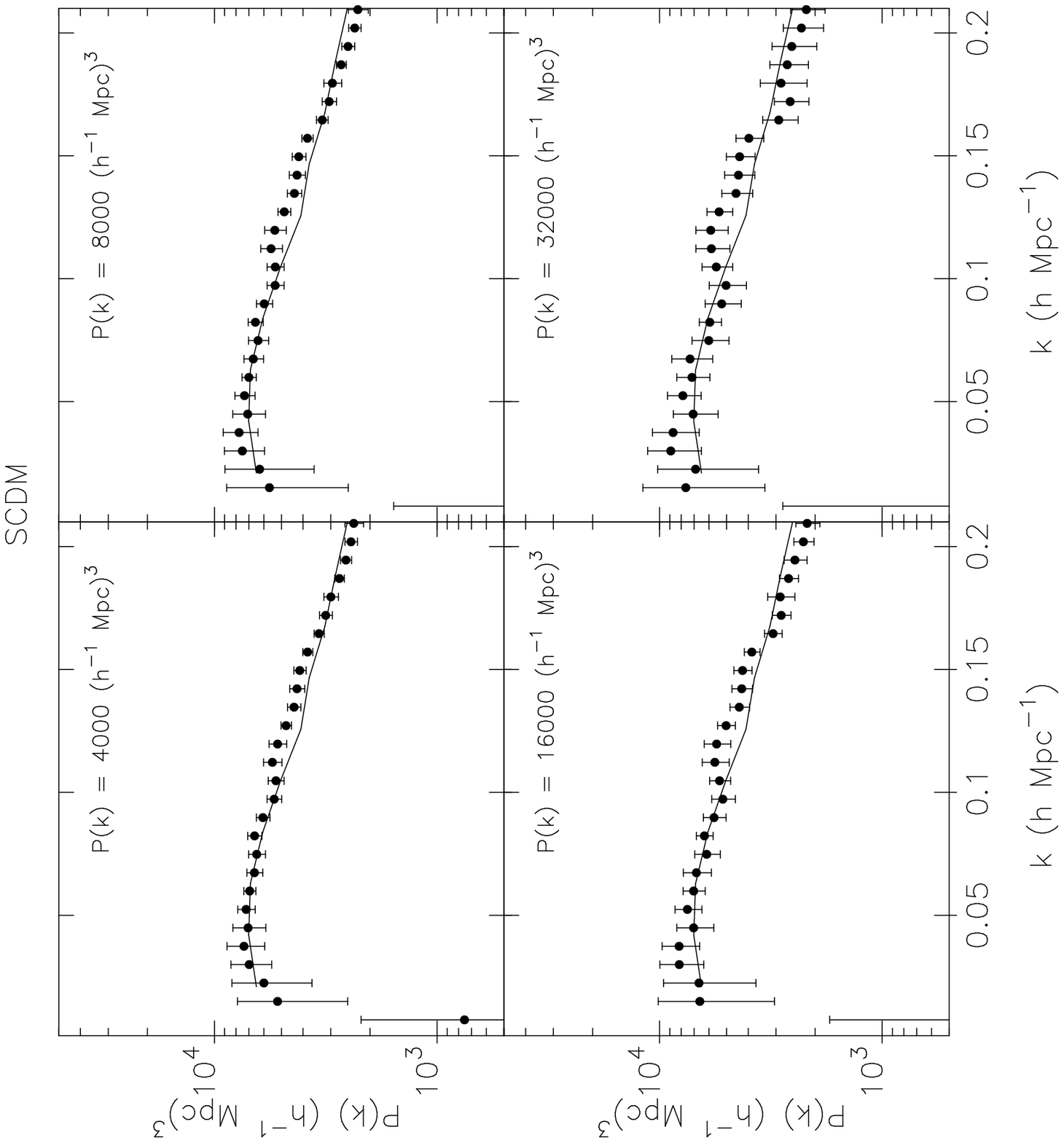}
\includegraphics{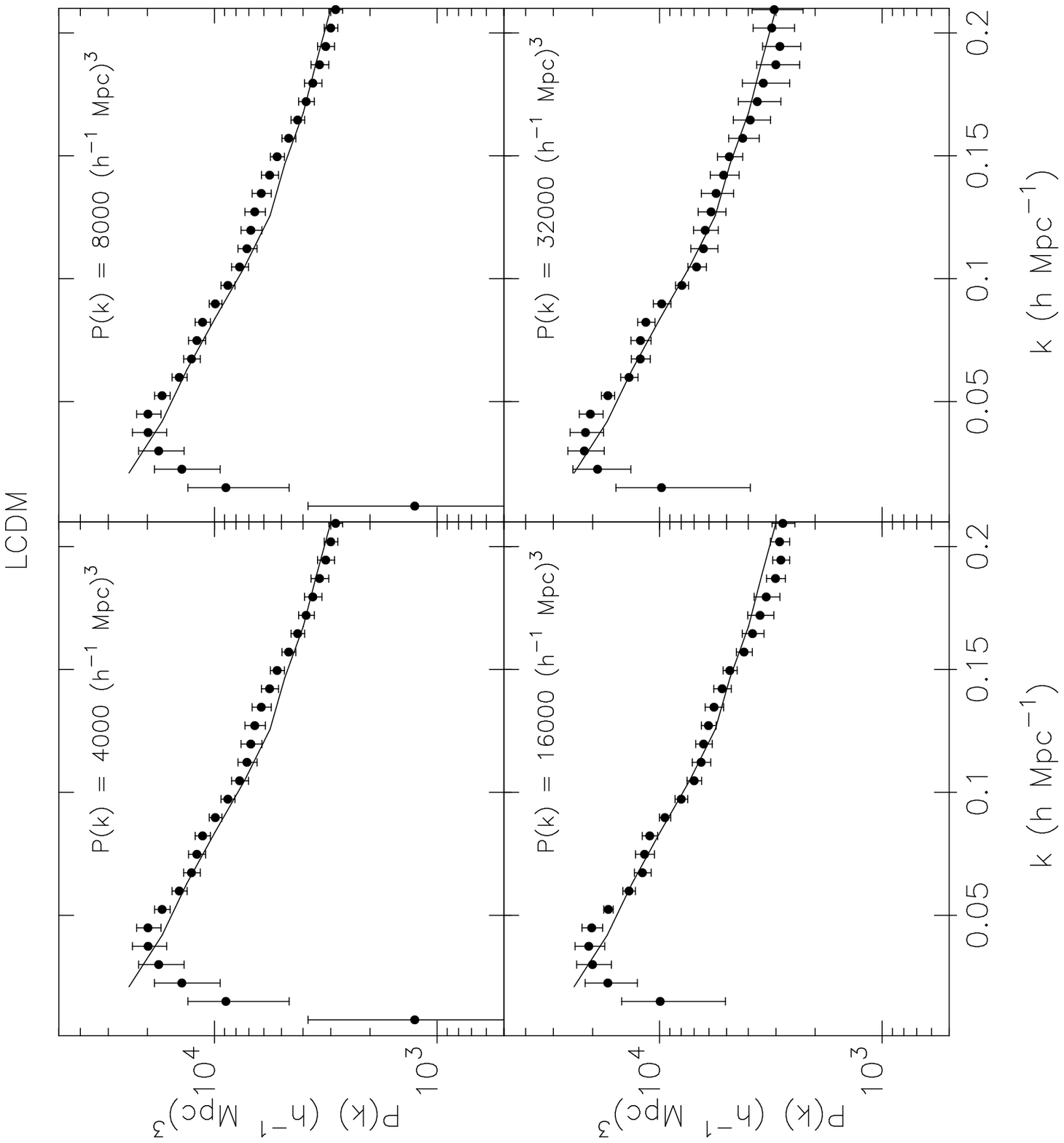}
\end{picture}
\caption{\label{stromps} Flux weighted power spectra determined
from mock sparsely sampled Stromlo-APM surveys. The upper figure
shows results for the SCDM ensemble and the lower figure shows
results for the LCDM ensemble. The solid line in each panel shows
the mean power spectrum determined from the full N-body simulations,
as plotted in Figure 3. The points show the average over ten mock
Stromlo-APM surveys for four values of $P(k)$ in the weighting scheme
of equation (15), $P(k) = 4000$, $8000$, $16000$, $32000(h^{-3}{\rm
Mpc})^3$, as indicated in each panel. The error bars show one standard
deviation on the mean.}
\end{figure}

\section{The Power Spectrum of the Stromlo-APM Survey}

\subsection{Comparison of results from volume limited and flux limited
samples}

In Figure~\ref{stromps} we show the power spectra for three volume
limited subsets of the Stromlo-APM survey, together with $1\sigma$
errors derived from the FKP formula (equation 14). The power spectra
are consistent with each other, with no obvious dependence on the
volume limit.  The results are qualitatively similar to those of the
simulations plotted in Figure 3. The power spectrum of the real survey
has an approximately power law behaviour $P(k) \propto k^{-1.7}$ at
wavenumbers $k \simgt 0.05 h {\rm Mpc}^{-1}$, flattens off and
declines at smaller wavenumbers. The decline is almost certainly caused
by the bias discussed in Appendix A and so is not a real feature of
the galaxy distribution. The Stromlo-APM survey contains little
information at wavenumbers $k \simlt 0.05 h {\rm Mpc}^{-1}$,
since these scales are comparable to or greater than the size of 
the survey.

\begin{figure}
\centering
\begin{picture}(400,250)
\includegraphics{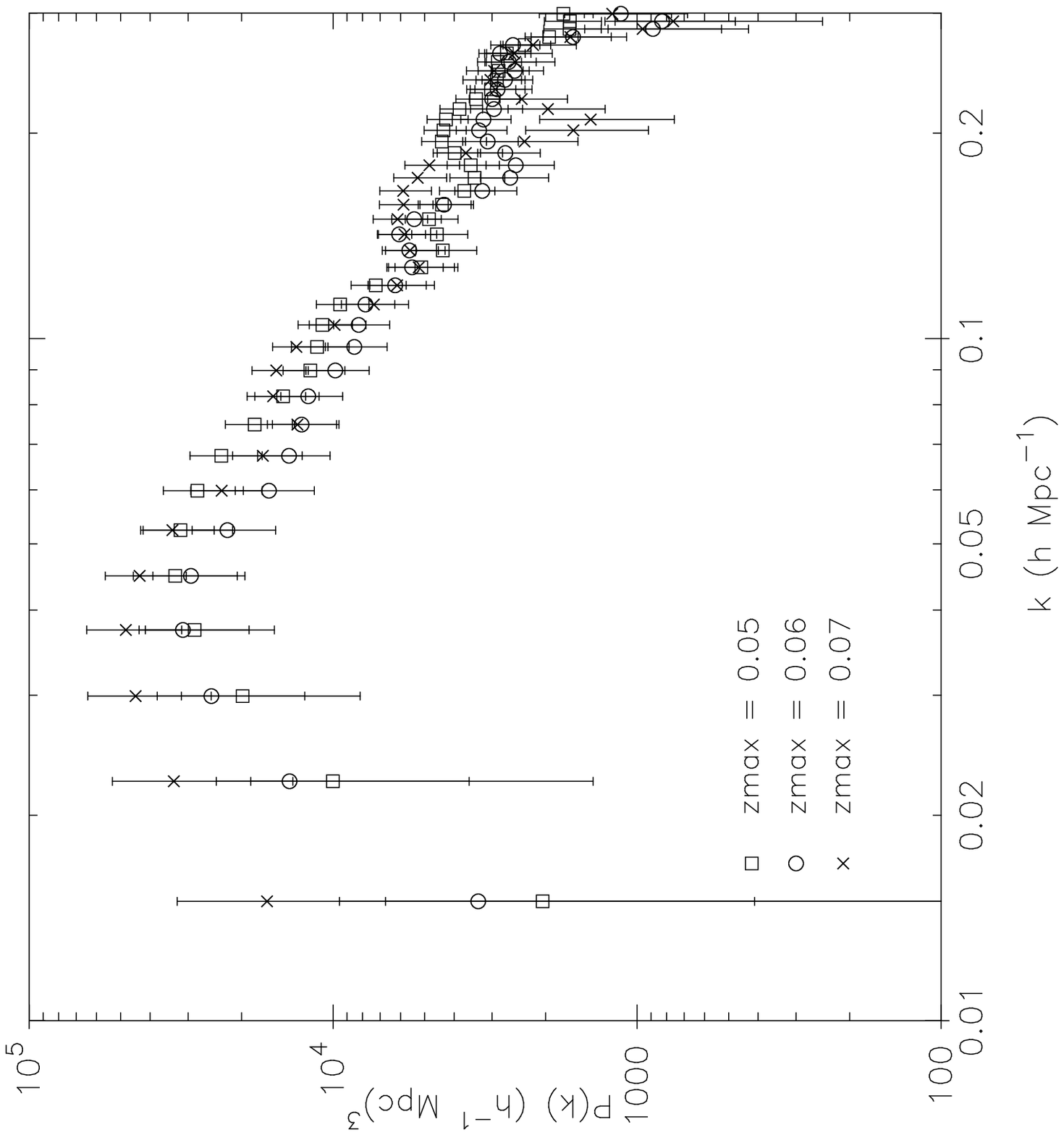}
\end{picture}
\caption{\label{stromps} Volume limited power
spectra of the Stromlo-APM survey, volume limited at $z =
0.05, 0.06, 0.07$ (150$\hmpc$, 180$\hmpc$, and 210$\hmpc$). }
\end{figure}

Figure 6 shows the flux limited power spectrum from the Stromlo-APM
survey for four values of $P(k)$ in the weighting function of equation
(15). As in our analysis of the flux limited mock Stromlo-APM
simulations (Figure 4) we see that the power spectra in Figure 6 are
insensitive to the weighting scheme. The power spectra in Figures 5
and 6 peak at a maximum value of $\sim 30 000 (h^{-3}{\rm Mpc})^3$ at
a wavenumber of $\sim 0.05 h {\rm Mpc}^{-1}$ and decline to a value of
$\sim 4 000 (h^{-3}{\rm Mpc})^3$ at wavenumbers $\sim 0.15 h {\rm
Mpc}^{-1}$. Thus, over most of the wavenumber range plotted in Figure
6, a value of $P(k) \sim 10^4(h^{-3}{\rm Mpc})^3$ in equation (15) 
should result in an estimate of $P(k)$ that is close to the
minimum variance estimate. In the remainder of this paper,
unless otherwise stated,  we will
use the flux limited power spectrum with $P(k)$ set to 
$8 000 (h^{-3}{\rm Mpc})^3$ in equation (15).

\begin{figure}
\centering
\begin{picture}(400,250)
\includegraphics{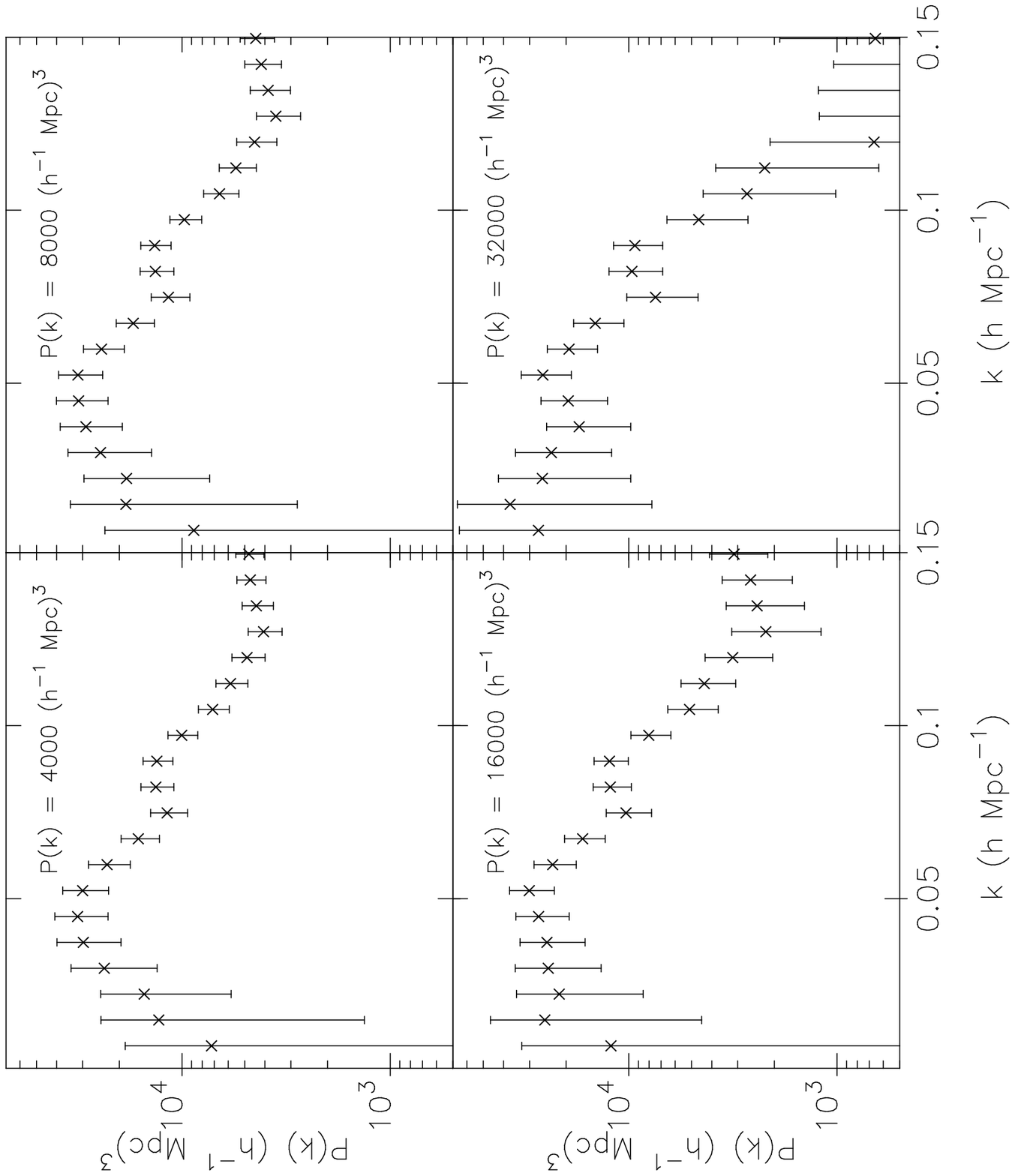}
\end{picture}
\caption{\label{compothers} Flux limited power
spectra of the Stromlo-APM survey for four values of 
$P\left(k\right)$ in the weighting scheme of equation 15,
$P\left(k\right)= 4000$, $8000$, $16000$, and $32000(h^{-3}{\rm
Mpc})^3$ as indicated in each panel.}
\end{figure}

In Figure 7, we compare the volume limited power spectrum estimate for
$z_{max} = 0.06$ with the flux limited estimate derived with $P(k) = 8
000 (h^{-3}{\rm Mpc})^3$ in equation (15).  The estimates are
consistent with each other within the $1 \sigma$ error bars. This
figure, and the results shown in figures 4 to 6, show that the power
spectrum estimates from the Stromlo-APM survey are robust. In summary,
we find no obvious differences between the power spectra of the volume
limited samples plotted in figure 5, or between the power spectra
measured from flux limited samples with different weighting functions.
This places some constraints on variations of the power spectrum as a
function of galaxy luminosity which we discuss in further detail in
the next subsection.

\begin{figure}
\centering
\begin{picture}(400,230)
\includegraphics{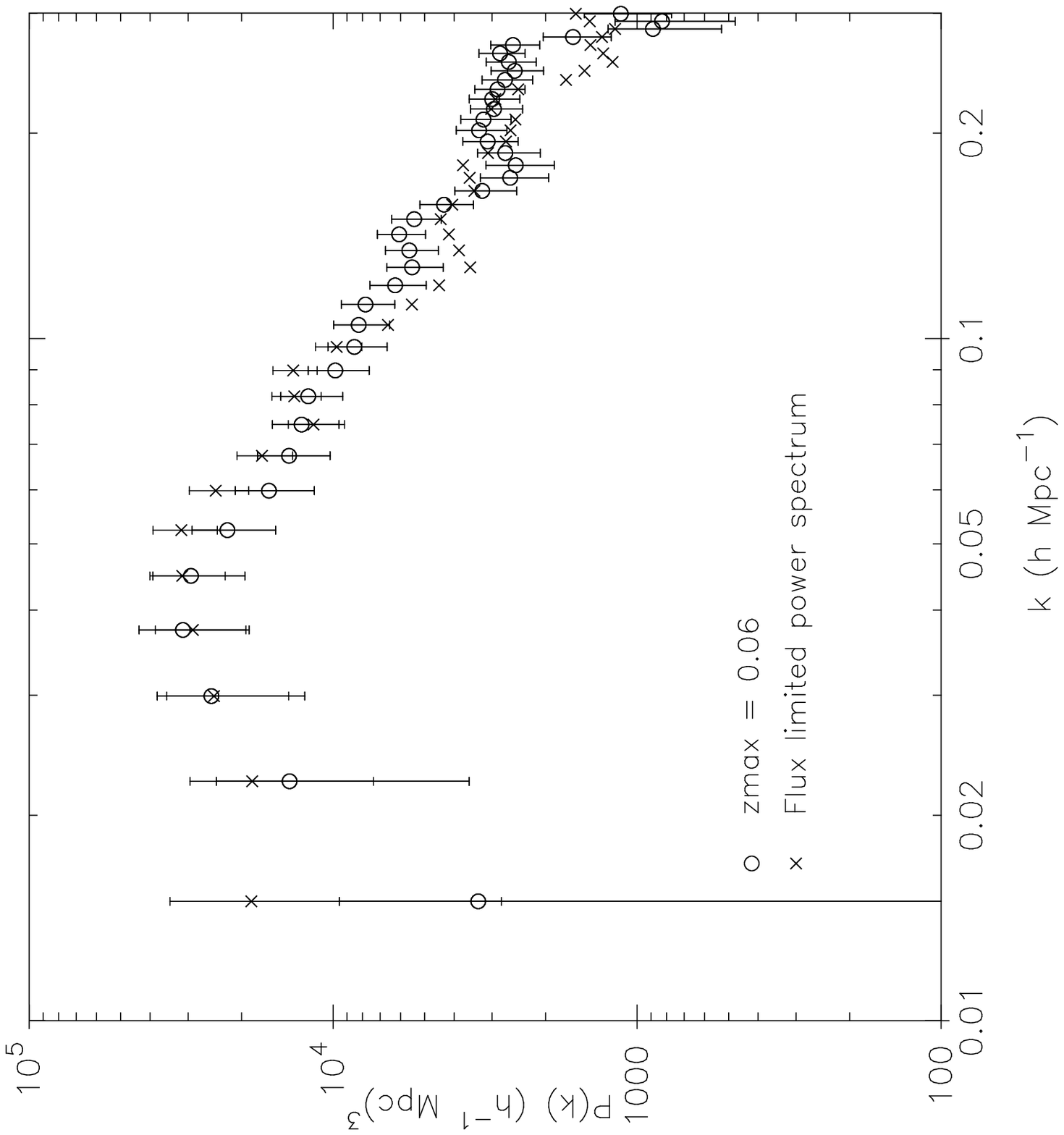}
\end{picture}
\caption{Comparison of the volume and flux limited power
spectra estimated from the Stromlo-APM survey.
The circles show the volume limited power
spectrum for the $z_{max}=0.06$ sample as plotted in Figure 5
and the crosses show the flux limited power spectrum
with $P(k) = 8 000 (h^{-3}{\rm Mpc})^3$ in equation (15)
as plotted in Figure 6. We plot $1\sigma$ error bars, but we have
suppressed the errors on the flux limited estimates at high
wavenumbers for clarity. }
\end{figure}

\subsection{Variations of the clustering strength with galaxy
luminosity}

As mentioned in the introduction, several groups have claimed to find
various correlations between the strength of the clustering pattern and
galaxy luminosity. In this section we investigate this possibility by
comparing the power spectrum measured from volume limited subsamples
of the Stromlo-APM data as a function of volume limit. We have already
seen from Figure 5 that there is little evidence for any systematic
dependence on the amplitude of $P(k)$ with volume limit over the
range $z_{max} = 0.05$--$0.07$, corresponding to cuts in absolute
magnitude of $M_{crit} = -18.91$ at $z_{max}=0.05$ and
$M_{crit} = -19.71$ at $z_{max} = 0.07$
(equation 1). As we increase the volume
limit beyond $z_{max} = 0.07$, the number of galaxies in the
Stromlo-APM survey declines exponentially and hence the errors on
$P(k)$ become large. We have therefore averaged the power spectrum
estimates over a range of wavenumbers to test for a linear bias
in $P(k)$ ({\it i.e} a change in amplitude independent of
wavenumber) as a function of volume limit. This has been done as
follows: we have computed the power spectra of six volume limited
samples with $z_{max} = 0.04$, $0.05$, $0.06$, $0.07$, $0.08$, $0.09$.
These power spectra were averaged to produce a mean power spectrum
$P_{av}(k)$. For each volume limit, we computed the ratio
$P(k)/P_{av}(k)$, which we averaged over the wavenumber range
$0.052 < k < 0.14 h {\rm Mpc}^{-1}$ to produce an estimate of
the relative bias factor $b^2(M_{crit})$ for each volume limit.
As can be seen from Figure 6, the error bars on $P(k)$ depend
weakly on wavenumber, hence it is reasonable to compute
$b^2(M_{crit})$ as an unweighted average over wavenumber.
These bias factors are plotted against $M_{crit}$ in Figure
8.  If there were no dependence of the clustering strength with
luminosity, we would expect the resulting bias factors to be close
to unity for all values of $M_{crit}$. In fact, we find that the
relative bias factor is close to unity for most of the magnitude
range, but is lower than unity for the faintest magnitude
cut at $M_{crit} = -18.4$ and higher than unity for the brightest
magnitude cut at $M_{crit} = -20.3$.

To assess the statistical significance of these result, we have
computed linear bias factors in exactly the same way for each of the
$30$ mock Stromlo-APM surveys constructed from the LCDM ensemble. The
results are plotted as the crosses in Figure 8 together with the
standard deviation for a single Stromlo-APM survey. The results from
the mock surveys are close to unity over the entire magnitude range,
as expected since there is no luminosity dependence of the clustering
pattern in the simulations. These results show that there is no firm
evidence for any luminosity dependence of the amplitude of the power
spectrum determined from the Stromlo-APM survey. At the faintest
magnitude cut, $7$ out of $30$ simulated catalogues gave a relative
bias factor lower than that measured for the real data. For the
brightest magnitude cut, only $2$ out of $30$ simulations gave a
relative bias factor larger than that determined from the Stromlo-APM
survey. Thus, there is tentative evidence that the amplitude of the
power-spectrum may increase at high luminosities. These results seem
consistent with previous work. Loveday \et (1995a) found some
evidence that the spatial two-point correlation function for
galaxies in the Stromlo-APM survey in the magnitude range  
$-19 < M < -15$ has a lower amplitude than that measured for
brighter galaxies. The results of Figure 8 suggest that this
may be caused by sampling fluctuations rather than a
real luminosity dependence of the clustering pattern. Park
\et (1994) have analyzed the power spectra of volume limited
subsets of the CfA-2 redshift survey and find evidence that
the amplitude of the power spectrum for galaxies brighter
than $M^*$ is about $40\%$ higher than the amplitude measured
for fainter galaxies. This is
consistent with the results plotted in Figure 8 and 
suggests that the rise in the relative bias factor
at $M_{crit} < - 20.3$ may be a real feature of the
galaxy distribution. Evidently, larger redshift surveys
are required to firmly establish whether this effect is
real but the results of this section show that over
most of the luminosity range, the power spectra measured
from the Stromlo-APM survey are consistent with the
null hypothesis that the clustering strength is independent
of luminosity.

\begin{figure}
\centering
\begin{picture}(200,200)
\includegraphics{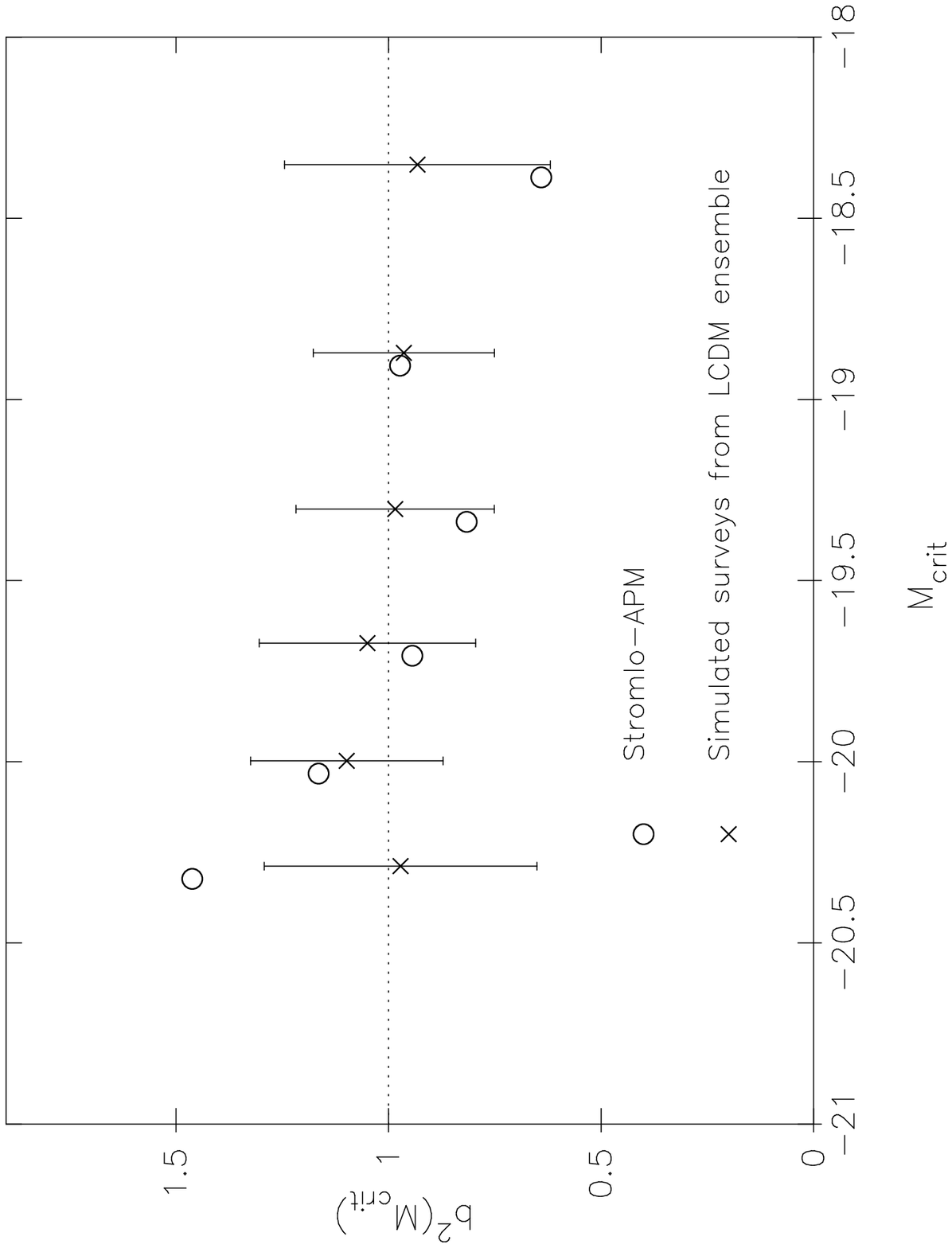}
\end{picture}
\caption{\label{lumbias} The relative bias factor $b^2$ defined
in the text plotted against limiting absolute magnitude $M_{crit}$.
Circles show the result for the Stromlo-APM data, crosses show
the results for simulated Stromlo-APM surveys drawn from
the LCDM ensemble. The error bars on these points show the standard
deviation of one simulation.}
\end{figure}

\section{Redshift-Space Distortions of the Power Spectrum}

Galaxy peculiar velocities cause a distortion of the clustering
pattern measured in redshift space compared to the true
pattern in real space (see {\it e.g.} Kaiser 1987). In this section,
we analyse the distortions to the shape of the power spectrum
measured in redshift space using the N-body simulations described
in Section 2. This allows us to quantify the effects of linear
and non-linear peculiar velocities and hence to  establish the range in
wavenumber over which linear perturbation theory can be used to
model the distortions. (For a  similar analysis of redshift-space
distortions in  N-body  simulations, see Gramman \et 1993). We then
compare the redshift-space estimates of $P(k)$ for the Stromlo-APM
survey of the previous section with the real-space estimates
of $P(k)$ for the APM Galaxy Survey derived by Baugh and Efstathiou
(1993, hereafter BE) to quantify the effects of redshift-space
distortion on large spatial scales.

\subsection{Redshift-space distortions of $P(k)$ determined from
N-body simulations}

Figure 9 shows the power-spectra measured in real-space and
redshift space for each of the ensembles of N-body simulations
described in Section 2. In each case, the amplitude of the
power spectrum measured in redshift space is enhanced 
on large scales compared to the power spectrum measured in real 
space.  In linear perturbation theory, the power
spectrum in redshift-space, $P_s(k)$, is related to the 
power-spectrum in real-space, $P_r(k)$, according to the
formula
\begin{equation}
P_{s}\left({\bf k}\right)=P_{r}\left({\bf
k}\right)\left(1+\beta\mu^{2}\right)^{2}, \qquad 
\beta = \frac{\Omega^{0.6}}{b},
\end{equation}
(Kaiser 1987). In this equation,  $\mu$ is the angle between the 
vector ${\bf{k}}$ and the line-of-sight, $b$ is a linear bias
factor relating fluctuations in the galaxy distribution to
fluctuations in the mass distribution, $ (\delta \rho /\rho)_{gal}
= b (\delta \rho/\rho)_{mass}$, and we have assumed that power
spectrum is determined from a patch of the universe that subtends
a small solid angle at the position of the observer.
Averaging equation (18) over the angle $\mu$,  we obtain
\begin{equation}
P_{s}\left(k\right)=P_{r}\left(k\right)\left(1 + 
\frac{2}{3}\beta +\frac{1}{5}\beta^2 \right).
\end{equation}
Thus the ratio $P_s(k)/P_r(k)$ depends on the parameter
$\beta = \Omega^{0.6}/b$ in linear perturbation theory\footnote
{This is true even for low density spatially flat models,
see Peebles (1984) and  Lahav \et (1991)}.

On small scales (wavenumbers $k \simgt 0.1 h {\rm Mpc}^{-1}$), 
the redshift-space  power spectra in Figure 9 are
suppressed in amplitude compared to the real-space spectra. 
This is caused by the small-scale peculiar velocities in
high density regions that produce so called  `fingers of God'
in redshift space. The solid lines in Figure 9 show a simple model
derived by Peacock and Dodds (1994) that incorporates both the
linear theory distortion of equation (18) with a model for the
distortions caused by fingers of God. The small scale peculiar
velocities are assumed to be uncorrelated in position and drawn
from a Gaussian distribution with 1-dimensional dispersion 
$\sigma_v$. The combined effects lead to the following
formula relating the angle averaged power spectra 
in redshift space and real space:
\begin{equation}
P_{s}\left(k\right) = P_{r}\left(k\right)G\left(\beta,k\sigma_v\right)
\end{equation}
where the function $G$ is given by
\begin{eqnarray*}
G(\beta, k \sigma_v)  = \frac{\sqrt{\pi}}{8}
\frac{erf\left(k\sigma_v\right)}{\left(k\sigma_v\right)^{5}}\left[3\beta^{2}
+ 4\beta \left(k\sigma_v\right)^{2} + 4\left(k\sigma_v\right)^{4}\right]
\end{eqnarray*}
\begin{eqnarray}
\qquad -
\frac{exp\left(-\left(k\sigma\right)^{2}\right)}
{4\left(k\sigma\right)^{4}}\left[
\beta^{2}\left(3 + 2\left(k\sigma\right)^{2}\right) +
4\beta \left(k\sigma\right)^{2}\right].
\end{eqnarray}
The solid lines in Figure 9 show equation (20) for each ensemble,
where we have used the 1-dimensional velocity dispersion
determined from the simulations $\sigma_v = 670.5$, $344.1$, and
$510.0$  $\kms$ for the SCDM, LCDM and MDM ensembles respectively.
These curves provide a good match to the results from the  N-body
simulations over the range of wavenumbers plotted
in Figure 9 and are more accurate than the model described
by Gramman \et (1993). Nevertheless, the model of equation (20) is simplistic, as it ignores non-linear streaming of galaxy pairs
(cf. eg. Nusser and Fisher (1995)).

\begin{figure}
\centering
\begin{picture}(300,350)
\includegraphics{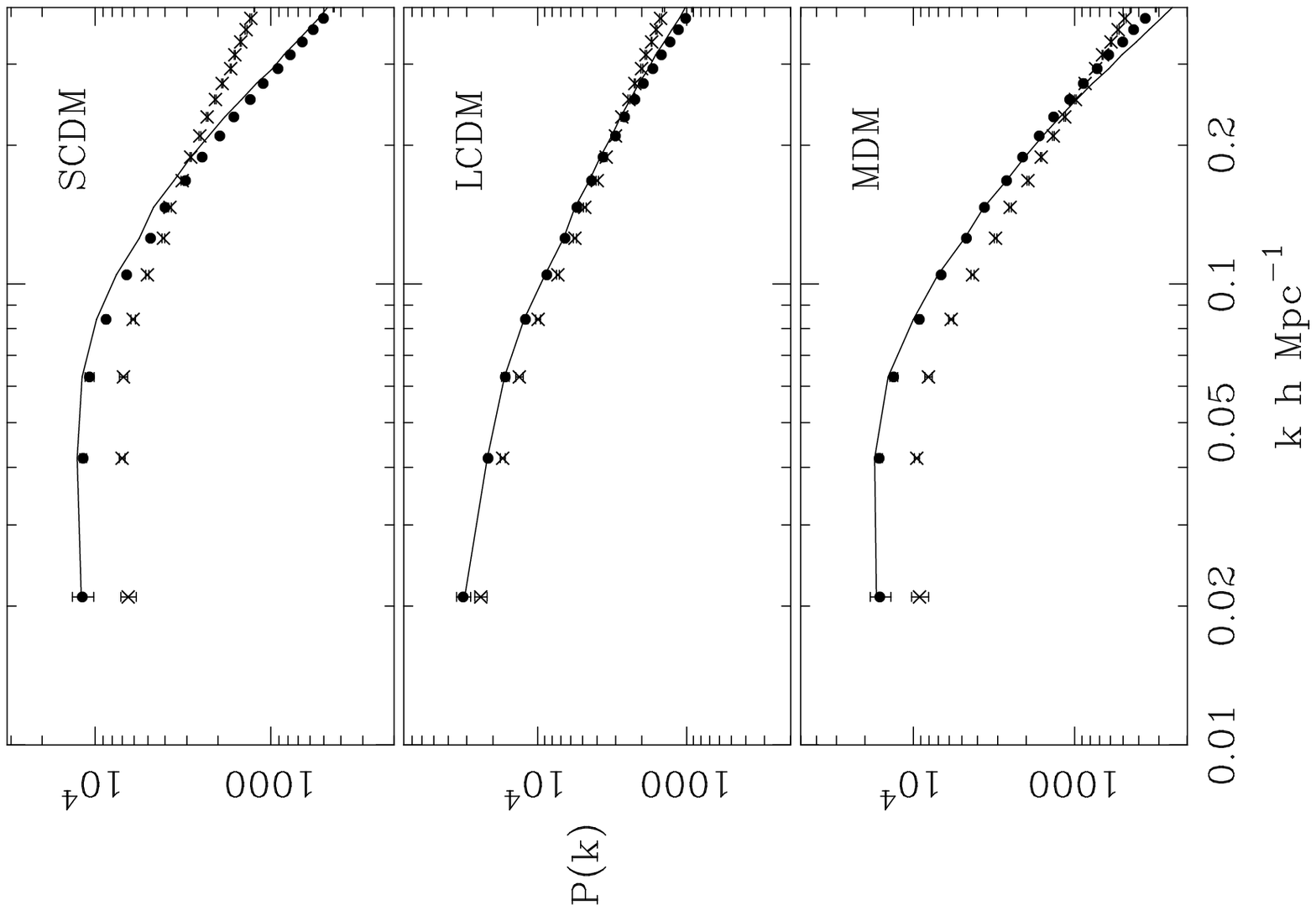}
\end{picture}
\caption{\label{plotdistfulln} Power spectra
evaluated in real and redshift space for each ensemble. Redshift space
power spectra are shown by the circles and real space spectra are shown by
the crosses. The solid lines show the predicted redshift space power 
spectrum using equation 20 and the measured real space power spectrum.
Values for $\sigma_v$ are given in the text.}
\end{figure}

Figure 10 shows the ratio of the power spectra plotted in Figure 9.
The solid lines show the linear theory relation of equation (19) 
and the dotted lines show equation (20). 
These results show that at wavenumbers $k \simgt 0.1 h {\rm Mpc}^{-1}$,
small-scale peculiar velocities cause a significant reduction in the 
amplitude of the power spectrum measured in redshift space. One must
therefore be cautious in  determining $\beta$ from redshift
distortions using the linear relations equations (18) and 
(19). The range of validity of linear theory
is strongly dependent on the amplitude of the small-scale
peculiar velocity field, as expected from equation (20).
The biases can be large, for example, even at $k = 0.1 h {\rm
Mpc}^{-1}$, the ratio of the power spectra in the MDM model
is equal to $1.57$ compared to the linear theory prediction 
of $1.87$. 

The amplitude of the small-scale peculiar velocity field is
relatively poorly constrained by observations. For example,
Davis and Peebles (1983) found  an {\it rms}
relative  peculiar velocity of $v_{12} =  340 \pm 40 {\rm km}{\rm s}^{-1}$
between galaxy pairs separated by $\sim 1 \hmpc$ from an analysis
of the CfA-1 survey and  Bean \et
1983 found $v_{12} = 250 \pm 50 {\rm km}{\rm s}^{-1}$ from an
analysis of a smaller, but deeper survey. However, values as high
as $600$--$800 {\rm km}{\rm s}^{-1}$ have been derived in the
literature (Hale-Sutton \et 1989, Mo \et 1993). The best
constraints on $v_{12}$ for optical galaxies come from an analysis
of the CfA2 and Southern Sky Redshift Survey by Marzke \et (1995)
who find $v_{12} =  540 \pm 180 {\rm km}{\rm s}^{-1}$. The error
on $v_{12}$ is large and the Marzke \et results are consistent with
the small-scale peculiar velocities ($\sigma_v \sim v_{12}/\sqrt{2}$)
for the mass points in the LCDM and MDM models but are lower than those of COBE
normalized SCDM models. In principle, with a larger survey, 
we could use equations
(20) and (21) (or a more complicated non-linear model) to simultaneously determine the parameters $\beta$ and
$\sigma_v$ (see Cole \et 1995, for an application to redshift
surveys of IRAS galaxies). 

\begin{figure}
\centering
\begin{picture}(300,350)
\includegraphics{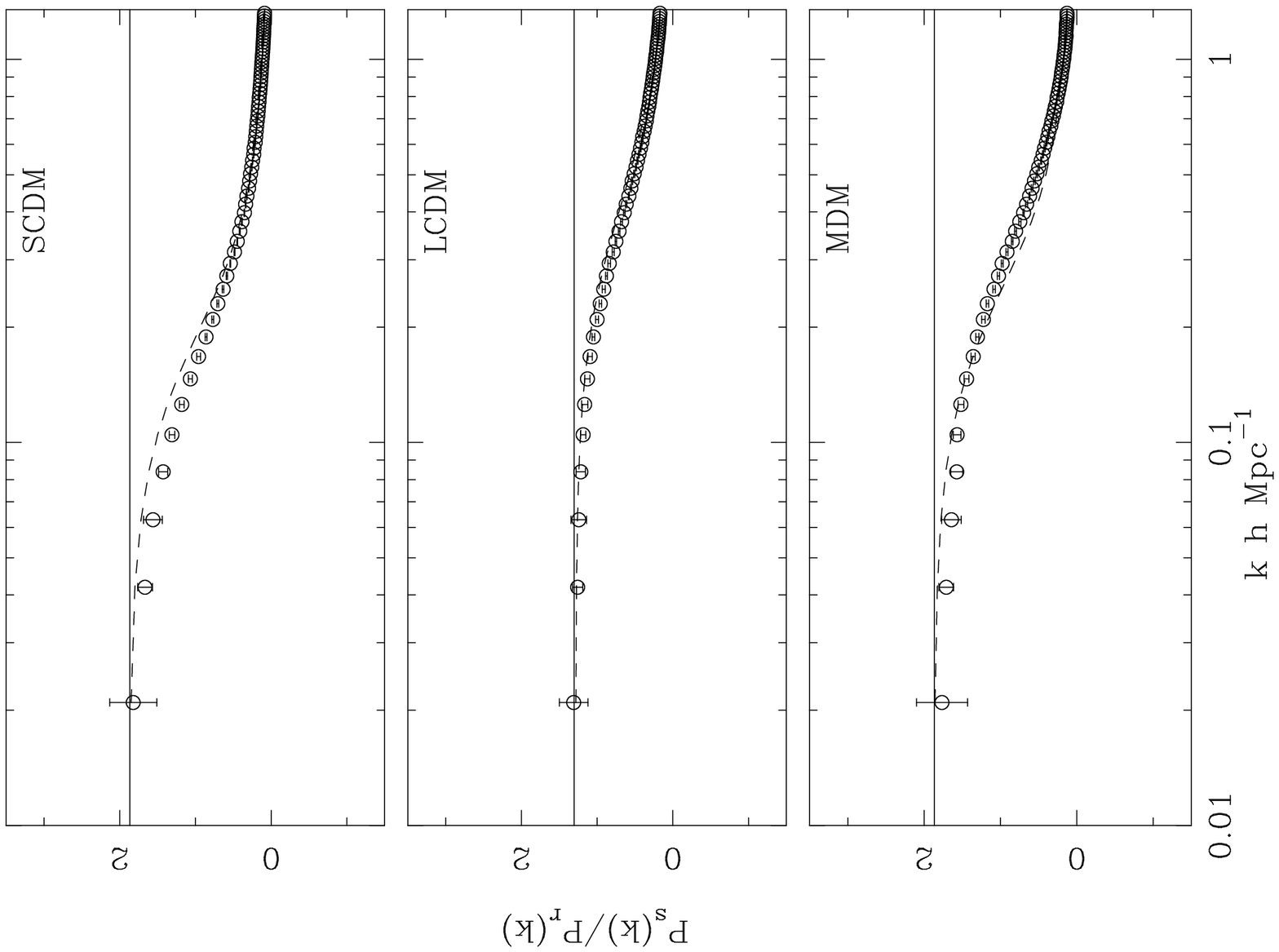}
\end{picture}
\caption{\label{plotfullratio} The ratio of redshift-space to
real-space power for the full N-body simulations (box size $300\hmpc$).
The solid lines shows the linear theory predictions and the dashed lines
show equation (20) using the same values of $\sigma_v$ as in Figure 9.} 
\end{figure}

\begin{figure}
\centering
\begin{picture}(300,350)
\includegraphics{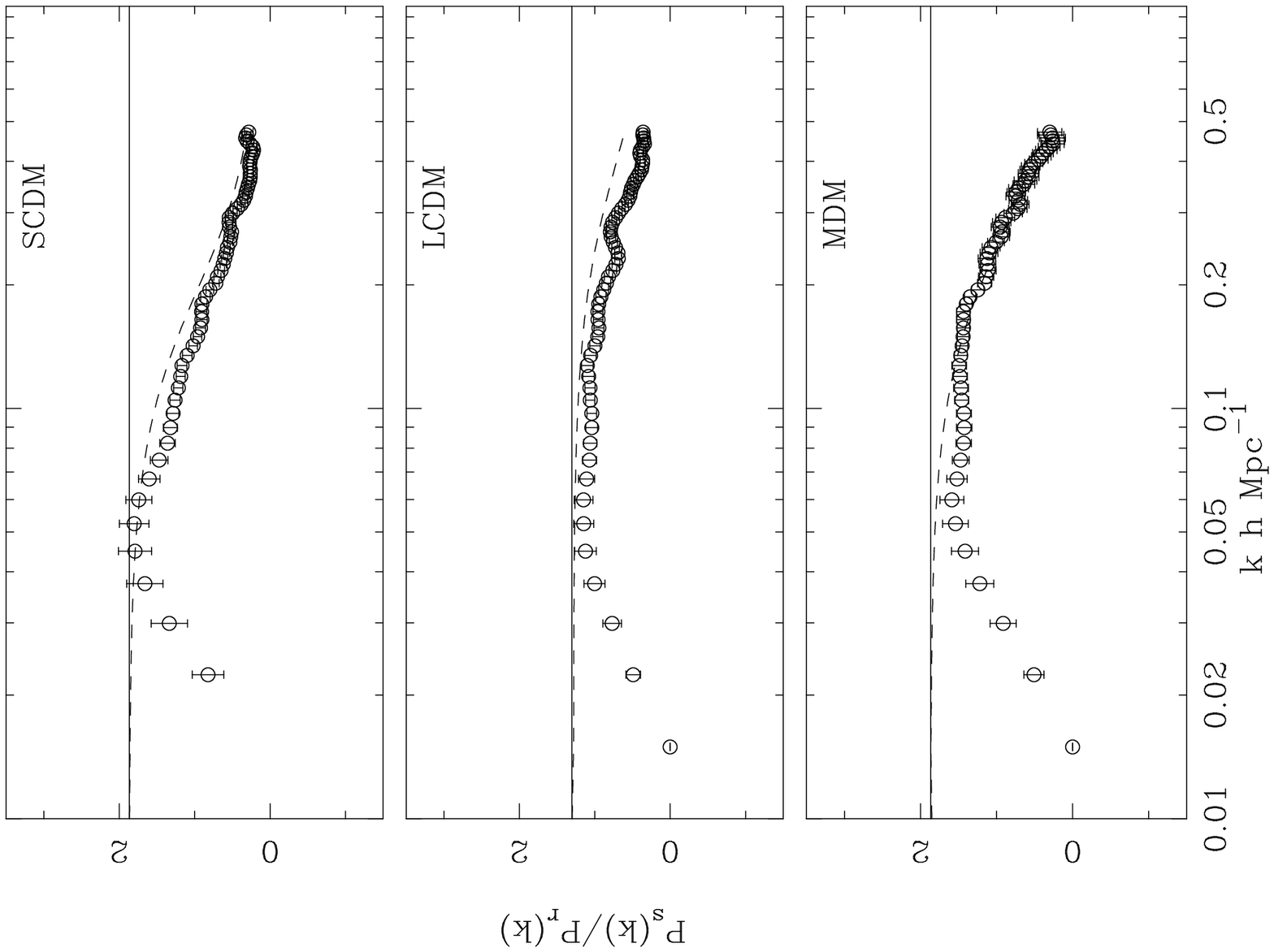}
\end{picture}
\caption{\label{plotsparserat} The ratios of redshift-space power 
spectra measured for sparsely sampled mock Stromlo-APM surveys 
to the real-space power spectra for the full N-body simulations.
The solid and dashed lines are as in Figure 10.}
\end{figure}

In addition to the effects of non-linear peculiar velocities,
biases in the estimates of $P(k)$ derived from equation (12) 
will also introduce departures from the linear theory 
predictions for the ratio $P_s(k)/P_r(k)$ at small
wavenumbers. This is illustrated in Figure 11, where we have
plotted the ratio of the redshift-space power spectra of the
mock Stromlo-APM simulations to the real-space power spectra
measured from the full N-body simulations. The decline in
the ratio at $k \simlt 0.05 h {\rm Mpc}^{-1}$ is caused by the
biases in the power spectrum estimates discussed in Appendix A.
Even at wavenumber $k \sim 0.05 h {\rm Mpc}^{-1}$,
the ratios for the LCDM and MDM ensembles fail to reach the
linear predictions primarily because the convolution of the 
redshift space power spectrum with the window function 
causes a slight depression in its amplitude ({\it cf} the
dotted lines in Figure 3).

\subsection{Estimates of $\beta$ from redshift space distortion}

In this Section, we estimate $\beta$ by comparing 
the Stromlo-APM power spectrum with real space estimates
of $P(k)$ for the APM survey. Baugh and Efstathiou (1993)
 have described a method for recovering the three-dimensional power
spectrum from the angular  two-point correlation function
$w(\theta)$.  The inversion requires a knowledge of the redshift
distribution of the galaxies used to estimate $w(\theta)$, of the
geometry of the Universe and of the evolution of the power spectrum.
The redshift distribution of APM galaxies is well constrained by
observations (see Maddox \et 1995). However, BE show that
uncertainties in the evolution of $P(k)$ and the cosmological
model introduce uncertainties of $\sim 20 \%$ in the amplitude of the
real space power spectrum.  As an illustration of these uncertainties, we
will use two estimates of the real space power spectrum derived from
APM galaxies in the magnitude range $17 < b_J < 20$ (plotted in Figure
12). In these inversions we assume an Einstein-de Sitter universe and
a power spectrum that evolves as $P(k,z) = P(k)/(1+z)^\alpha$ with
$\alpha = 0$ (filled circles) and $\alpha = 1.3$ (open circles)
corresponding to a clustering pattern that is stable in comoving and
physical coordinates respectively (see BE for further details).
Uncertainties in the redshift distribution and cosmological model
introduce similar systematic differences in the real space power
spectra, but since these errors are considerably smaller than the random
errors in the redshift space estimates of $P(k)$ , we will not discuss
them in detail here.

\begin{figure}
\centering
\begin{picture}(200,200)
\includegraphics{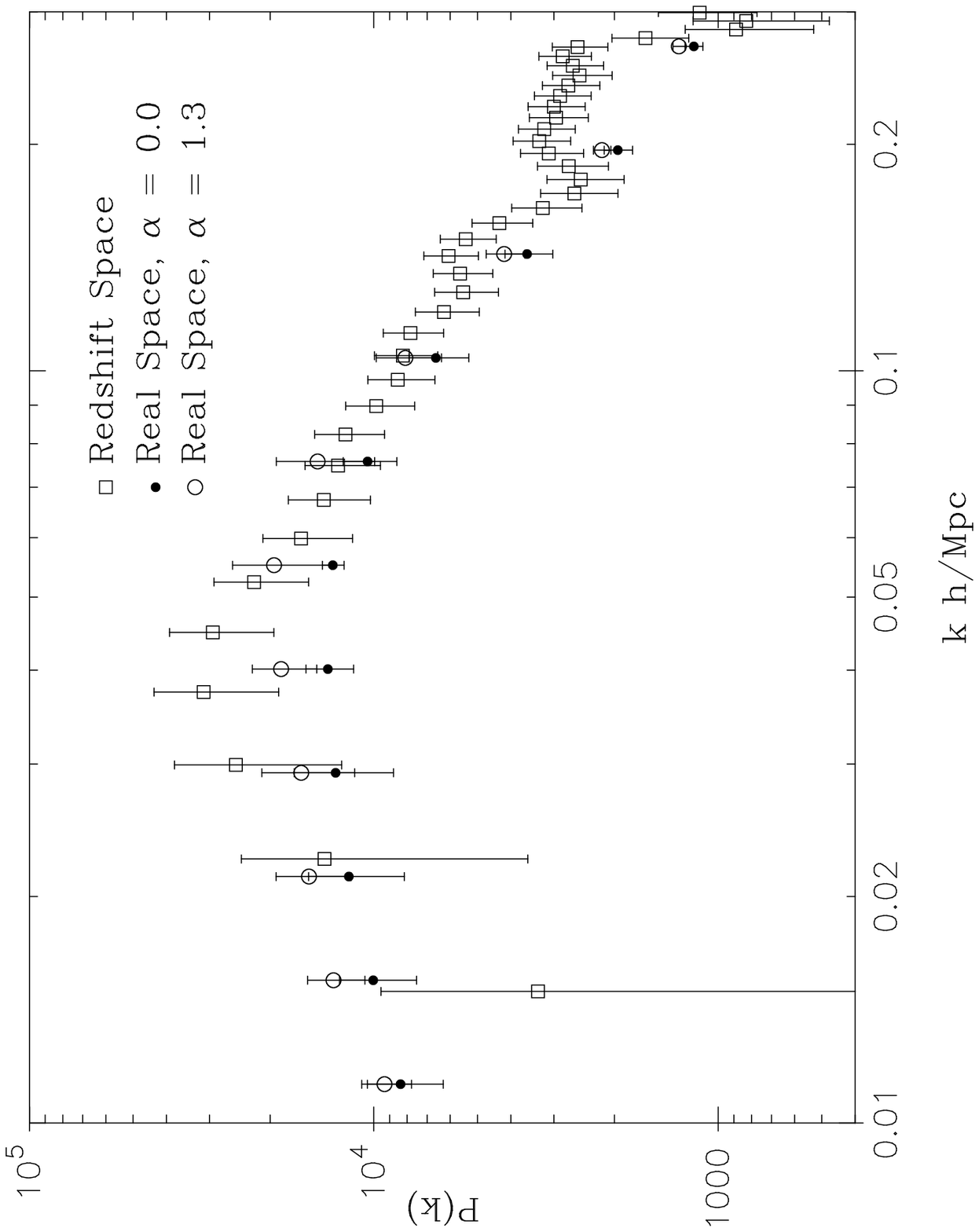}
\end{picture}
\caption{\label{compcmb}The open squares  show
the redshift-space power spectrum for a volume limited sample
from the Stromlo-APM survey limited at $z_{max} = 0.06$. The
circles show real space power spectra determined
using the methods of BE to invert the angular correlation function
of APM galaxies in the magnitude slice $17 < b_J < 20$. The filled
and open circles show the inversion assuming $\alpha = 0$ and
$\alpha = 1.3$ respectively, where the power spectrum is assumed
to evolve according to $P(k, z) = P(k)/(1+z)^\alpha$. We have plotted
$1\sigma$ errors on the Stromlo-APM estimates, as in Figure 7. The
error bars on the real space power spectra show the standard deviation
determined from the scatter in the estimates derived
from 4 roughly equal area zones of the APM survey as discussed in BE.}
\end{figure}

The open squares in Figure~\ref{compcmb} show the $z_{max} = 0.06$
redshift-space estimate of $P(k)$ for the Stromlo-APM survey as
calculated in Section 2.2. Qualitatively, the behaviour is similar to
that seen in the simulations; at wavenumber $k \simlt 0.03
\;h{\rm Mpc^{-1}}$ the redshift-space power spectrum declines and 
falls below the real space estimates. Over the wavenumber range
 $0.03 \simlt k \simlt 0.07\;h{\rm Mpc^{-1}}$ the redshift-space amplitude is
enhanced suggesting a significant distortion of the clustering
pattern in redshift space. At larger wavenumbers, the errors in
the Stromlo-APM power spectrum become large because it is a 
sparse sampled redshift survey and so contains little information
on the small-scale clustering of galaxies.

We estimate the amplitude of the redshift space distortions in the
Stromlo-APM survey as follows.  We compute the ratio $P_s(k)/P_r(k)$
by linear interpolating the real-space power spectra plotted in Figure
12 and we average this ratio over the wavenumber range $0.05 < k < 0.1
\;h{\rm Mpc^{-1}}$. This leads to an estimate of $\beta$ from equation
19. To estimate the error in $\beta$, we compute the scatter in the
ratio of the redshift-space to real-space power-spectra from the mock
Stromlo-APM surveys plotted in Figure 11 averaged over the same
wavenumber range as the data. This provides an estimate of errors
associated with sampling fluctuations in the redshift survey. We estimate the
error associated with uncertainties in the real space estimate of
$P(k)$ from the scatter in the ratio $P_s(k)/P_r(k)$ using $P_r(k)$
determined from four roughly equal area zones of the APM survey
(see BE). Our
final error estimate on $\beta$ is derived by adding these two error
terms in quadrature on the assumption that they are uncorrelated.
Table 1 shows our estimates of $\beta$ derived using the two
real-space estimates of $P(k)$ plotted in Figure 12, and two estimates
of the redshift-space power spectrum for the Stromlo-APM survey; the
$z_{max} = 0.06$ volume limited results plotted in Figure 12 and the
flux-limited results from Figure 6, with a weight function (equation
15) in which we have set $P(k) = 8000 (h^{-1} {\rm Mpc})^3$.

\begin{table}
\centering
\begin{tabular}{||c|c|c||}
\hline \hline 
     & $\alpha$=0.0 & $\alpha$=1.3   \\ \hline
     & $\beta$  & $\beta$   \\ \hline
Flux Weighted $P_s(k)$  & 0.74 $\pm$ 0.48 & 0.20 $\pm$ 0.44   \\
Volume Limited $P_s(k)$ & 0.38$\pm$ 0.67 & -0.13 $\pm$ 0.64   \\ \hline
\end{tabular}
\caption{\label{tablesim} Estimates of $\beta$ from the Stromlo-APM survey.}
\end{table}

These results suggest a positive value of $\beta$, though the errors
are large. The differences in the flux-weighted and volume limited
estimates of $P(k)$ over the wavenumber range $0.05 < k < 0.1 \;h{\rm
Mpc^{-1}}$ lead to an uncertainty of nearly a factor of two in the
derived value of $\beta$. A similar uncertainty in introduced by the
difference in the real-space estimates of $P(k)$ derived from the APM
survey for the two adopted values of the evolution parameter $\alpha$.

The inferred errors on $\beta$ are larger than those quoted in other
analyses of redshift space distortions. For example, Hamilton (1993)
finds $\beta = 0.66^{+0.34}_{-0.22}$ from an analysis of angular
moments of the two-point correlation function of the $1.9$Jy IRAS
redshift survey of Strauss \et (1992); Cole \et (1995) analyse angular
harmonics of the power spectrum and find $\beta = 0.52 \pm 0.13$ and
$\beta=0.54 \pm 0.3$ for the 1.2Jy and QDOT IRAS surveys respectively.
Loveday \et (1995b) find $\beta = 0.48 \pm 0.12$ from an analysis of
anistropies in the redshift-space correlation function of the
Stromlo-APM survey. At face value our estimates of $\beta$ appear to
disagree with those of Loveday \et, who analyse the same survey.
However, it seems likely that Loveday \et have underestimated the
error on $\beta$. As a further check of whether our errors are
realistic, Table 2 list the errors and biases in $\beta$ derived
by applying the above analysis to the mock Stromlo-APM surveys
constructed from the simulations.  Here we have computed $\beta$ from
the ratio of the redshift-space power spectra of the mock surveys to
the mean real-space power spectrum for each ensemble, and we have
neglected errors in the real-space estimate $P(k)$ since these are
negligible. In each case, the errors in $\beta$ are close to $\delta
\beta \approx 0.5$ in good agreement with our error estimates for the
real data. In addition, Table 2 shows that $\beta$ is underestimated
in each case because of the nonlinear corrections described in the
previous subsection and because of the biases in the redshift-space
estimates of $P(k)$ described in Appendix A.

The results given in Table 1 are in good agreement with those
determined by Baugh (1995) from a comparison of the redshift-space 
two-point correlation function of the Stromlo-APM survey and
the real-space correlation function inferred by inverting 
the angular correlation function of the APM survey.

Our analysis suggests that the results of Table 1 should be considered
as lower bounds on the true value of $\beta$ and hence that $\beta =
1$ is compatible with the observations. With larger redshift surveys
it should be possible to fit a more complicated model to the data so
extending the useable range of wavenumbers and reducing the biases in
$\beta$, {\it e.g.} equation 21 with $\beta$ and $\sigma_v$ as free
parameters, or the nonlinear model of Nusser and Fisher (1995) based
on the Zeldovich approximation.

\begin{table}
\centering
\begin{tabular}{||c|c|c|c||}
\hline \hline 
Model & $\beta$ & $\delta\beta$  & Linear $\beta$ \\ \hline
SCDM  & 0.64 & 0.54 & 1.00  \\
LCDM  & 0.13 & 0.55 & 0.38  \\ 
MDM   & 0.62 & 0.52 & 1.00   \\ \hline
\end{tabular}
\caption{\label{tablesim} Estimates of $\beta$ derived from the ratio
$P_s/P_r$ over the wavenumber range $0.05 < k < 0.1
\;h{\rm Mpc^{-1}}$ from mock Stromlo-APM surveys.}
\end{table}

\section{Conclusions}

The conclusions of this paper are as follows:

\noindent
[1] We have tested estimators of $P(k)$ using simulated redshift
surveys constructed from N-body simulations. These tests show that
estimates of $P(k)$ are biased at low wavenumbers if the surveys are
used to estimate the mean galaxy density. Formulae for these
biases are given in Appendix A.

\noindent
[2] We have estimated $P(k)$ for the Stromlo-APM redshift survey using
volume limited and flux limited samples. The power spectra are
insensitive to the volume limit and to the galaxy weights applied in
the analysis of flux limited samples.

\noindent
[3] We have investigated whether the amplitude of $P(k)$ depends on
galaxy luminosity and used N-body simulations to assess the
statistical significance of our analysis. We find no evidence for any
significant luminosity dependence except possibly at absolute
magnitudes brighter than $M = -20.3$, where we find some evidence for
a higher amplitude. This is broadly  consistent with the analysis of
the CfA-2 redshift survey by Park \et (1994) who find that the
amplitude of $P(k)$ for galaxies brighter than $M^*$ is about
$40\%$ higher than the amplitude measured from fainter galaxies.

\noindent
[4] We have analysed numerical simulations to determine the effects of
redshift-space distortions on the shape of $P(k)$.  The distortions
can be well approximated by the formula of Peacock and Dodds (1994),
equation (20), which depends on two parameters, $\beta =
\Omega^{0.6}/b$ and a measure of the small-scale {rms} peculiar
velocity, $\sigma_v$.

\noindent
[5] We estimate the redshift-space distortions in the Stromlo-APM
survey by comparing  redshift-space power spectra with estimates
of the real-space power spectrum determined by inverting the angular
correlation function measured for the parent APM Galaxy Survey. The
results indicate a positive value of $\beta$, consistent with other
work, but the uncertainties are large and do not exclude $\beta=1$.

\begin{figure}
\centering
\begin{picture}(400,450)
\includegraphics{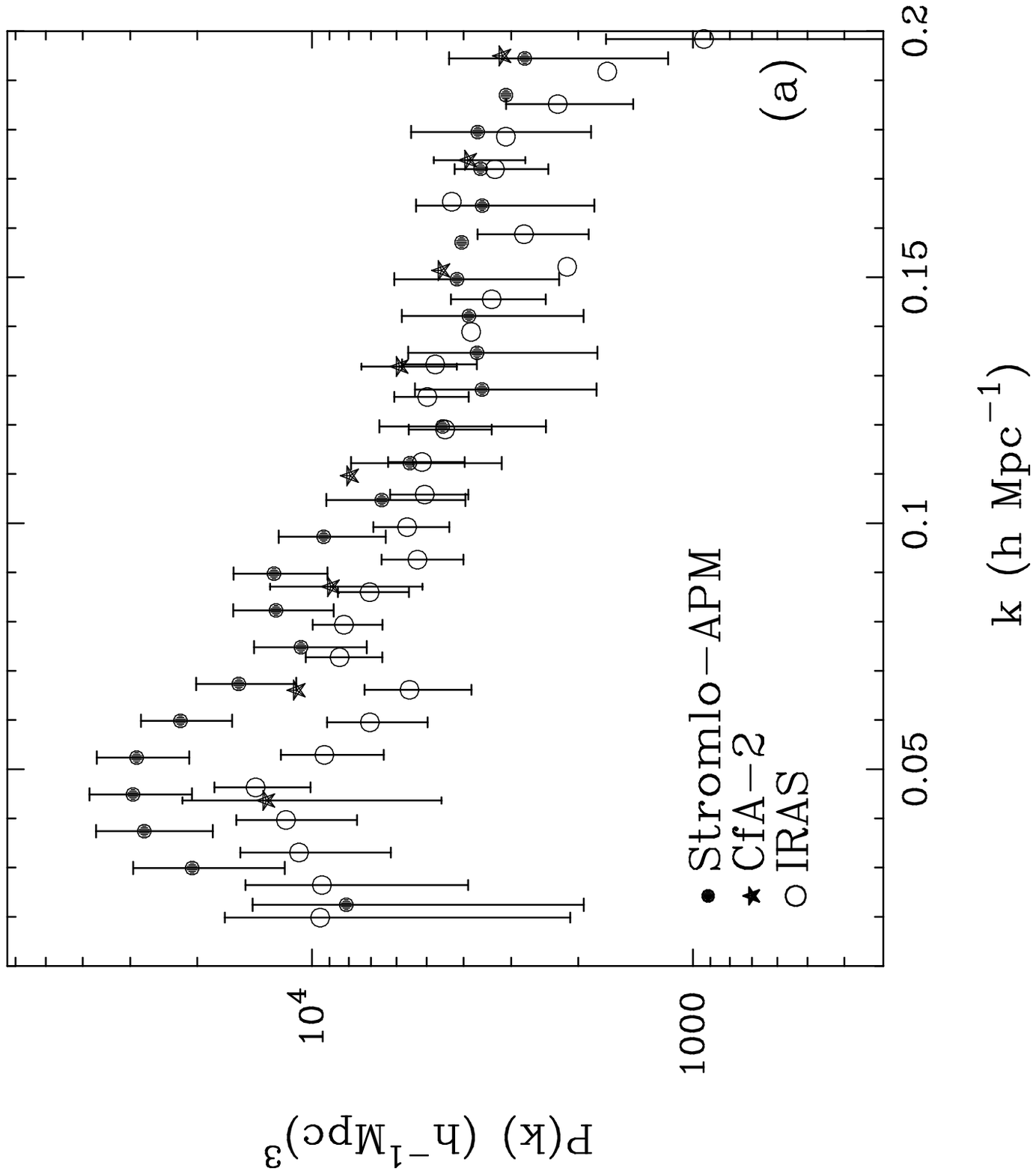}
\includegraphics{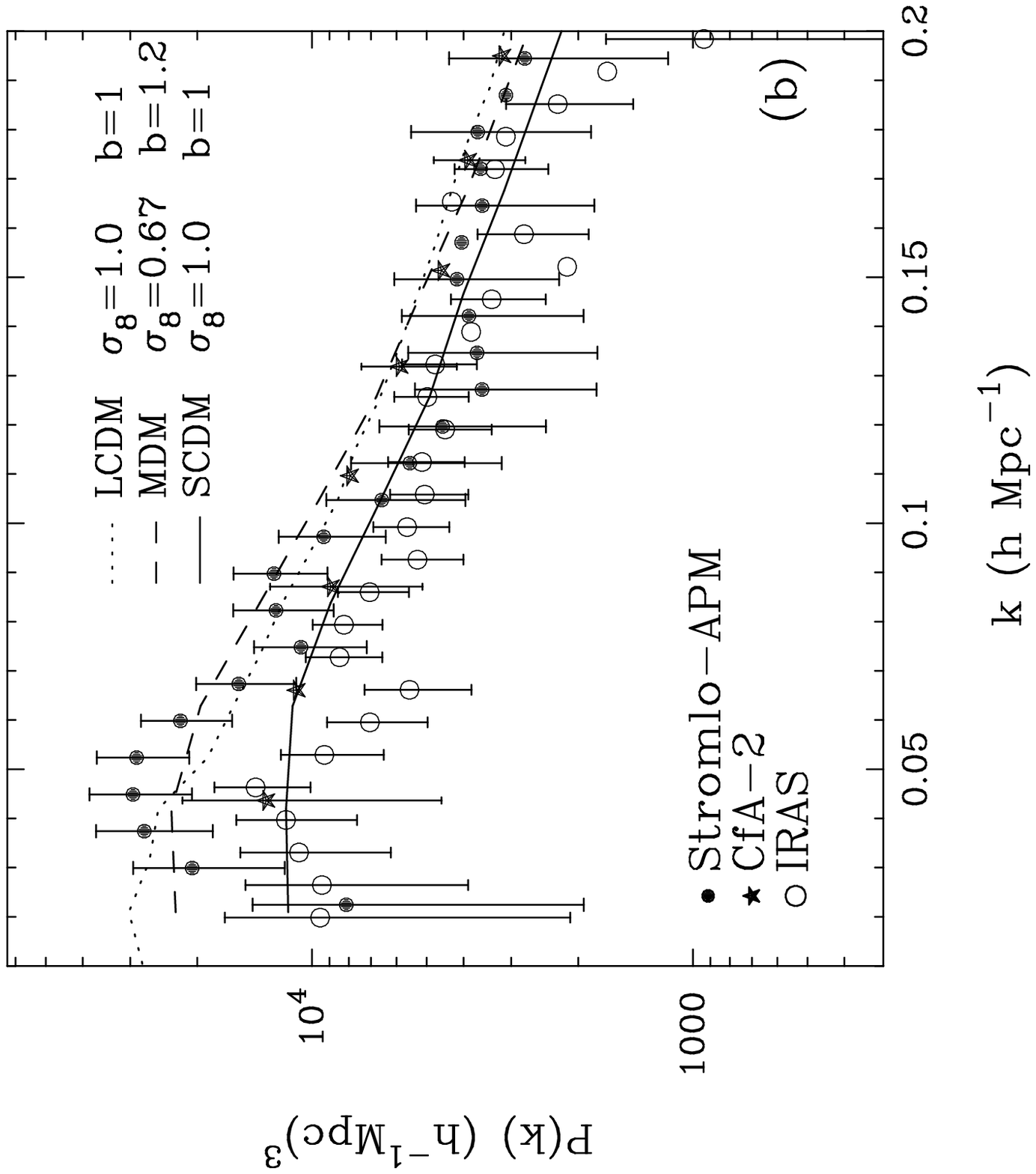}
\end{picture}
\caption{\label{pspec} Estimates of the power spectra for various
redshift surveys. The filled circles show our flux-limited estimates
of $P(k)$ for the Stromlo-APM survey, as plotted in Figure 7. The open
circles show a flux limited estimate of $P(k)$ for IRAS galaxies from
Tadros and Efstathiou (1995). The filled stars show estimates of
$P(k)$ from Park \et (1994) for a volume limited subset of the CfA-2
survey. The three curves in Figure (13b) show redshift space power
spectra determined from the mass distributions in N-body simulations
of the three CDM-like models described in the text; $\sigma_8$ gives
the {\it rms} amplitude of the mass fluctuations in spheres of radius
$8 h^{-1}{\rm Mpc}$ and we have multiplied the power spectrum of the
MDM model by $b^2 = (1.2)^2$ to approximately match the power-spectra
of the Stromlo-APM and CfA-2 surveys.}
\end{figure}

In Figure 13, we compare the Stromlo-APM power spectrum with power
spectra estimated from other surveys. The filled circles show the flux
limited estimates of $P(k)$ for the Stromlo-APM survey using $P(k) =
8000 \left(\hmpc\right)^{3}$ in the weighting scheme of equation (15).
The open symbols show $P(k)$ for the combined $1.2$Jy and QDOT IRAS
redshift surveys, as analysed by Tadros and Efstathiou (1995); these
estimates are for a flux limited sample with $P(k) = 8000
\left(\hmpc\right)^{3}$ in the weighting scheme of equation (15). The
filled stars show $P(k)$ from Park \et (1994) for a volume limited
subset of the CfA-2 survey consisting of $1509$ galaxies to a
coordinate distance of $101 h^{-1} Mpc$. The power spectra for the
two optically selected catalogues are consistent with each other
within the errors, but the amplitude of the IRAS power spectrum
is slightly lower indicating that IRAS galaxies are less strongly
clustered than optically selected galaxies. The results shown in
Figure 13 are consistent with a linear relative bias between optical
and IRAS galaxies of $b_{opt}/b_{IRAS} \sim 1.2$. Recently, Landy \et (1996) have computed a 2 dimensional power spectrum for the Las Campanas redshift survey and find evidence for a strong peak in the power spectrum at wavenumbers $k \sim 0.06 h {\rm Mpc}^{-1}$. Their results are not easily comparable to ours and it is unclear whether there is any inconsistency with the three dimensional power spectrum inferred for the APM Galaxy survey (Baugh and Efstathiou 1993) or with the redshift space power spectrum of the Stromlo-APM survey presented here. 
A detailed comparison is deferred to a subsequent paper.
 
In Figure 13b, we show the empirical estimates of $P(k)$ together with
redshift-space estimates of $P(k)$ determined from N-body simulations
of the LCDM, MDM, and SCDM models. The shape of the SCM curve is in
good agreement with the power spectra measured for the IRAS and CfA-2
surveys, though it lies low compared to the Stromlo-APM data points at
wavenumber $k \sim 0.05 h {\rm Mpc}^{-1}$. The high amplitude for the
SCDM model implied by the COBE temperature anisotropies results in a
high amplitude for the peculiar velocity field. This in turn leads to
a significant difference in the shape of the redshift-space power
spectrum compared to the real-space power spectrum (see also Bahcall
\et 1993) and so the model can provide an acceptable match to the
redshift-space $P(k)$ although it fails to match the shape of the
real-space power spectrum (see {\it e.g.} BE, Maddox \et 1995). As
mentioned in Section 6.1, the small scale ($\sim 1 h^{-1} {\rm Mpc}$)
relative peculiar velocities predicted by the COBE normalised SCDM
model are considerably higher than the relative motions
between galaxy pairs measured by Marzke \et (1995). Both the LCDM
and MDM models provide acceptable fits to the power spectra measured
from the redshift surveys and distinguishing between these models
would require accurate estimates of the effects of redshift space
distortions. As we have seen from the analysis of the Stromlo-APM
survey,  larger redshift surveys,
such as the AAT 2-degree field redshift survey (see Efstathiou 1996) 
and the Sloan Digital Sky Survey (Gunn \& Weinberg 1995) are required
to determine $\beta$ accurately and so distinguish between these
models.

\section*{Acknowledgment}
HT and GPE acknowledge the receipt of a PPARC studentship and Senior
Fellowship respectively. We thank Carlton Baugh and Jon Loveday for
useful conversations and for comments on the manuscripts. We thank Jon
Loveday, Bruce Peterson and Steve Maddox for their invaluable
contributions to the Stromlo-APM redshift survey.

\bigskip
\centerline{\large References}
\parskip=0pt
\refindent
Bahcall N. A., Cen R. \& Gramman M., 1993, ApJ {\bf 408}, L77
\refindent
Baugh C.M. \& Efstathiou G., 1993, MNRAS, {\bf 265}, 145.
\refindent
Baugh C.M. \& Efstathiou G., 1994, MNRAS, {\bf 267}, 323.
\refindent
Baugh C. M., 1995 {\it in preparation}
\refindent
Baumgart D. J. \& Fry J. N., 1991, ApJ, {\bf 375}, 25
\refindent
Bean A. J., Efstathiou G., Ellis R. S., Peterson B. A. \& Shanks T.,
1983, MNRAS, {\bf 205}, 605
\refindent 
Bond J. R. \& Efstathiou G., 1984, ApJ, {\bf 285} L45
\refindent
Cole S., Fisher K. B. \& Weinberg D. H., 1995, MNRAS, {\bf 275}, 515
\refindent
Croft R. A. C., \& Efstathiou G., 1994, MNRAS, {\bf 268}, L23
\refindent
Davis M. \& Peebles P. J. E., 1983 Apj, {bf 267}, 465
\refindent
Davis M., Efstathiou G., Frenk C. S. \& White S. D. M., 1985, ApJ,
{\bf 292}, 371
\refindent
Efstathiou G., Davis M., Frenk C. S. \& White S. D. M., 1985, ApJS,
{\bf 57}, 241
\refindent
Efstathiou G., Ellis R. S. \& Peterson B. A., 1988, MNRAS, {\bf 232}, 431
\refindent
Efstathiou G., 1996, {\it `Observations of Large-Scale Structure
in the Universe'}, Les Houches Lectures 1995, ed R. Schaeffer,
Elsevier Science Publishers, Netherlands, in press.
\refindent
Feldman H.A., Kaiser N. \& Peacock J.A., 1994, ApJ,
{\bf 426}, 23.
\refindent
Fisher K.B., Davis M., Strauss M.A., Yahil A. \& Huchra J.P.,
1993, ApJ, {\bf 402}, 42.
\refindent
Gramman M., Cen R. \& Bahcall N. A., 1993, ApJ, {\bf 419}, 440
\refindent
Gunn J. E. \& Weinberg D. H., 1995, {\it `Wide Field Spectroscopy and
the Distant Universe, The $35^{th}$ Herstmonceux Conference'}
eds. S. J. Maddox \& A. Aragon-Salamanca
\refindent
Hale-Sutton D., Fong R., Metcalfe N. \& Shanks T., 1989, MNRAS, {\bf
237}, 569
\refindent
Hamilton A. J. S., 1988, ApJ, {\bf 331}, L59
\refindent
Hamilton A. J. S., 1993, ApJ, {\bf 406}, L47
\refindent 
Iovino A., Giovanelli R., Haynes M., Chincarini G. \& Guzzo L., 1993,
MNRAS, {\bf 265}, 21
\refindent
Kaiser N., 1987, MNRAS, {\bf 227}, 1
\refindent
Klypin A., Holtzman J., Primack J. \& Regos E., 1993, ApJ, {\bf 416},
1
\refindent
Lahav O., Lilje P. B., Primack J. R. \& Rees M., MNRAS 1991, {\bf
251}, 128
\refindent
Landy S. D., Schectman S. A., Lin H., Kirshner R. P., Oemler A. A. \& Tucker D., 1996 {\it {preprint}}
\refindent
Loveday J., Efstathiou G., Peterson B. A. \& Maddox S. J., 1992a, ApJ,
{\bf 400}, L43
\refindent
Loveday J., Peterson B. A., Efstathiou G. \& Maddox S. J., 1992b, ApJ,
{\bf 390}, 338
\refindent
Loveday J., Maddox S. J., Efstathiou G. \& Peterson B. A., 1995a, ApJ,
{\bf 442}, 457
\refindent
Loveday J., Efstathiou G., Maddox S. J. \& Peterson B. A., 1995b {\it preprint}
\refindent
Maddox S. J., Sutherland W. J., Efstathiou G. \& Loveday J., 1990a,
MNRAS {\bf 243}, 692
\refindent
Maddox S. J., Efstathiou G. \& Sutherland W. J., 1990b, MNRAS, {\bf
246},433
\refindent
Maddox S. J., Efstathiou G. \& Sutherland W. J., 1995 {\it submitted}
\refindent
Marzke R. O., Geller M. J., da Costa L. N. \& Huchra J. P., 1995 {\it
preprint} 
\refindent
Mo H. J., Jing Y. P. \& B\"orner G., 1993, MNRAS, {\bf 264}, 825
\refindent
Nusser A. \& Fisher K. B., 1995 {\it preprint}  astro-ph/9510049
\refindent
Park C., Gott J. R. \& da Costa L. N., 1992, ApJL, {\bf 392}, L51
\refindent
Park C., Vogeley M. S., Geller M. J., \& Huchra J. P., 1994 ApJ, {\bf
431}, 569
\refindent
Peacock J. A. \& Dodds S. J., 1994, MNRAS, {267}, 1020
\refindent
Peebles P. J. E., 1984 ApJ, {\bf 284}, 439
\refindent
Santiago B. X. \& da Costa L. N., 1990, ApJ, {\bf 362}, 386
\refindent
Schechter P. L., 1976, ApJ, {\bf 203}, 297 
\refindent 
Smoot G. F. \et 1992, ApJ, {\bf 396}, L1
\refindent
Strauss M. A., Huchra J. P., Davis M., Yahil A., Fisher K. B. \& Tonry
J., 1992, ApJS, {\bf 83}, 29
\refindent
Tadros H. \& Efstathiou G., 1995, MNRAS, {\bf 276}, L45
\refindent
Vogeley M. S., Park C., Geller M. J. \& Huchra J. P., 1992, ApJ, {\bf
391} L5

\vskip 0.5 truein

\noindent
{\bf Appendix A: Biases in estimating $P(k)$}

The quantity $\Delta(k)$ defined in equation (10) requires
an estimate of the mean galaxy density $\overline n$. Usually,
this estimate will be derived from the sample itself, {\it e.g.}
for a volume limited sample we can estimate $\overline n$ from
\begin{eqnarray*}
 \overline n = \frac{1}{V_s} \sum_i n_i W({\bf x}_i)  \qquad\qquad (A1)
\vspace{0.5cm}
\end{eqnarray*}
where $V_s$ is the volume of the survey and the sum extends over all
infinitesimal cells of galaxy count $n_i$. Inserting (A1) into
equation (10), we find that the expectation value of 
$\vert \Delta(k) \vert^2$ is given by,
\begin{eqnarray*}
\langle \vert \Delta ({\bf k}) \vert^2 \rangle
 = \frac{\overline n}{V} \sum_{\bf k^\prime} \vert \hat
W ({\bf k^\prime}) \vert^2 + \frac{\overline n^2}{V}
\sum_{\bf k^\prime} \vert \hat
W ({\bf k} - {\bf k^\prime}) \vert^2 P({\bf k^\prime}) \\
-\frac{\overline n}{V_s} \vert \hat W ({\bf k^\prime}) \vert^2
+ \frac{\overline n^2}{V_s^2}\int\int \xi(x_{12}) W({\bf x}_1)
W({\bf x}_2) \vert \hat W ({\bf k^\prime}) \vert^2 \;dV_1\;dV_2 \\
- \frac{\overline n^2}{VV_s}\int\int \xi(x_{12}) W({\bf x}_1)
W({\bf x}_2) \hat W ({\bf k^\prime}) {\rm e}^{i{\bf k \cdot x}_1}
\;dV_1\;dV_2 \\
- \frac{\overline n^2}{VV_s}\int\int \xi(x_{12}) W({\bf x}_1)
W({\bf x}_2) \hat W ({\bf k^\prime})^* {\rm e}^{-i{\bf k \cdot x}_2}
\;dV_1\;dV_2  \qquad (A2)
\end{eqnarray*}
The last four terms in equation (A2) account for the biases in
estimating $P(k)$ when the mean galaxy density is determined
from the sample itself. The first double integral in equation (A2)
measures the excess variance in the galaxy fluctuations
above Poisson noise averaged over the survey volume,
\begin{eqnarray*}
\sigma^2_s = \frac{1}{V_s^2} \int\int \xi(x_{12}) W({\bf x}_1)
W({\bf x}_2)  \;dV_1\;dV_2. \qquad \qquad (A3)
\end{eqnarray*}
If this variance is dominated by fluctuations on scales smaller than
the scale of the survey, it is a good approximation to set the other
two integrals in (A2) equal to equation (A3). An approximate
expression for the bias in $\vert \Delta(k) \vert^2$ is therefore,
\begin{eqnarray*}
-\frac{\overline n}{V_s} \vert \hat W ({\bf k^\prime}) \vert^2
- \overline n^2 \sigma^2_s \vert \hat W ({\bf k^\prime}) \vert^2
\qquad (A4.1) \\
\equiv 
-\frac{\overline n}{V_s} \vert \hat W ({\bf k^\prime}) \vert^2
- \overline n^2 \left ( \sum_{\bf k^\prime} 
\vert W ({\bf k^\prime}) \vert^2 P({\bf k^\prime}) \right)
 \vert \hat W ({\bf k^\prime}) \vert^2  (A4.2) \\
\end{eqnarray*}
where in (A4.2) we have written $\sigma^2_s$ in terms of the
power spectrum $P(k)$. We have used these equations to compute
the dashed lines shown in Figure 3 in Section 4.2. The first
term in each of equations (A4.1) and (A4.2) introduces a bias in the estimate
of $P(k)$ (equation 12) that depends on the number of galaxies in
the survey, but is independent of the shape of the power spectrum;
the second term introduces a bias that is independent of the 
number of galaxies in the sample, but depends on the shape of the
power spectrum.

It is straightforward to generalise this analysis to the FKP
estimator of $P(k)$ for flux limited samples. As discussed
in Section 3.1, we estimate the mean galaxy density by
computing the sum
\begin{eqnarray*}
\frac{1}{V_{eff}} \sum_i n_i w^\prime(x_i) W({\bf x}_i) \qquad (A5.1) \\
\end{eqnarray*}
where
\begin{eqnarray*}
V_{eff} = \int  w^\prime(x_i) W({\bf x}_i) \;dV \qquad (A5.2) \\
\end{eqnarray*}
and the weight function $w^\prime(x)$ is equal to 
\begin{eqnarray*}
w^\prime = \frac{1}{(1 + 4 \pi \overline n(x) J_3)}. \qquad (A5.3) \\
\end{eqnarray*}
Inserting this expression into equation (2.1.4) of FKP, we can
calculated the bias in a similar way to the calculation of equation
A2. As the final expression is lengthly, we do not reproduce it here.

\bibliographystyle{/data/castor/ht/strompaper/mnras}
\bibliography{/data/castor/ht/strompaper/general}
\end{document}